\documentclass{article}

\usepackage{PRIMEarxiv}

\usepackage[utf8]{inputenc} 
\usepackage[T1]{fontenc}    
\usepackage{hyperref}       
\usepackage{url}            
\usepackage{booktabs}       
\usepackage{amsfonts}       
\usepackage{nicefrac}       
\usepackage{microtype}      
\usepackage{lipsum}
\usepackage{appendix}
\usepackage{fancyhdr}       
\usepackage{graphicx}       

\usepackage{tabularx}
\usepackage{booktabs}
\usepackage{multirow}
\usepackage{makecell}
\usepackage{array}
\usepackage{hhline}
\usepackage{xcolor}
\usepackage{xcolor}
\usepackage[most]{tcolorbox}
\newenvironment{appendixlinedblock}{%
  \begin{tcolorbox}[
    breakable,
    colback=gray!6,
    colframe=gray!35,
    boxrule=0.45pt,
    arc=1mm,
    left=8pt,
    right=8pt,
    top=7pt,
    bottom=7pt,
    before skip=8pt,
    after skip=10pt
  ]
}{%
  \end{tcolorbox}
}
\newenvironment{agentdescription}[1]{%
    \par\medskip
    \begingroup
    \small
    \noindent
    \colorbox{black!8}{%
        \parbox{\dimexpr\linewidth-2\fboxsep\relax}{%
            \textbf{#1}%
        }%
    }\par
    \smallskip
    \noindent
    \ignorespaces
}{%
    \par
    \endgroup
    \medskip
}
\newenvironment{questionnaireitem}[1]{%
    \par\medskip
    \begingroup
    \small
    \noindent\rule{\linewidth}{0.8pt}\par
    \nopagebreak
    \noindent
    \colorbox{black!10}{%
        \parbox{\dimexpr\linewidth-2\fboxsep\relax}{%
            \centering\bfseries #1%
        }%
    }\par
    \nopagebreak
    \noindent\rule{\linewidth}{0.4pt}\par
    \nopagebreak
    \ignorespaces
}{%
    \par\smallskip
    \noindent\rule{\linewidth}{0.8pt}\par
    \endgroup
    \medskip
}

\newcommand{\qfield}[1]{%
    \par\smallskip\noindent\textbf{#1:} %
}

\newcommand{\placeholder}[1]{%
    \ensuremath{\langle}\textit{#1}\ensuremath{\rangle}%
}

\graphicspath{{media/}}     

\title{IMPACT: Modeling Socially Interdependent Movement in a Generative Multi-Agent Simulation of a Pompeian Household
}

\author{
\textbf{Tianqi Liu}$^{1}$ \quad
\textbf{Nayoung Kim}$^{1}$ \quad
\textbf{Julia Sebastien}$^{2}$ \\[0.4em]
\textbf{Andrea Stevenson Won}$^{2,\ast}$ \\
\\
$^{1}$Department of Information Science, Cornell University, Ithaca, NY 14853, USA \\
$^{2}$Department of Communication, Cornell University, Ithaca, NY 14853, USA
}

\begin{document}
\maketitle

\begin{abstract}
Simulations of archaeological sites can make interpretations of past cultural practices observable and examinable. Generative multi-agent simulations offer a bottom-up approach to modeling how people collectively moved through and used historical spaces. However, current agents designed to simulate everyday life often plan and act independently, limiting their ability to capture how movement depends on others' actions. We introduce IMPACT (Interdependent Movement Planning through Inter-Agent Constraints and Triggers), an architecture that uses culturally specific roles and obligations to define dependencies among agents' activities and guide coordination. IMPACT connects socially gated milestone planning, wait-or-prompt resolution, structured directive issuance, and directive integration. These mechanisms determine whether and when activities can begin or change as social conditions evolve, producing socially constrained and prompted movement as their primary observable outcome. We instantiate IMPACT in a five-hour simulation of a Pompeian dinner involving ten agents across interdependent roles. Analysis of five simulation runs shows how social roles, responsibilities, and status relations shape household activities and spatial practices, as reflected in patterns of co-location, asymmetric waiting, co-movement, and social directives. In a controlled ablation evaluation, thirty-seven participants rated the complete architecture's behavior as more socially coherent and believable than that of two reduced architectures. Interviews with six archaeology experts highlighted historically plausible movement patterns and the simulation's potential to support archaeological interpretation, while identifying areas requiring stronger historical grounding for future work. IMPACT demonstrates how agent-based modeling can operationalize culturally situated interpretations of past sociospatial practices.
\end{abstract}


\section{Introduction}

Agent-based simulation has long helped archaeologists study population movement, settlement, and human decision-making ~\cite{graham2022enchantment, barcelo2016simulating,cegielski2016rethinking,romanowska2021agent}. Previous reconstructions have populated historical environments with autonomous characters, demonstrating the potential of simulation to make past settings and social activities visible for archaeological research and public engagement~\cite{maim2007populating,antunes2022virtual}. Recent generative multi-agent systems (MASs) extend this approach by populating reconstructed environments with agents that form plans, act, and have dynamic social interactions over time~\cite{park2023generative,piao2026agentsociety}. However, to our knowledge, no simulation has yet examined how a person's movement is shaped by the presence, responsibilities, actions, and responses of other people in the group~\cite {kaiser2011roman,baines1990restricted}. 

Modeling collective use of space requires more than overlaying independent trajectories.
Practices such as hospitality, domestic service, dining, and ritual involve people with different social roles. Whether, when, and with whom a person moves may therefore depend on others' behavior, rather than be triggered simply by the passage of time or the completion of an individual task. We describe movement as \textit{socially constrained} when it depends on the presence or actions of relevant others. Movement is \textit{socially prompted} when an invitation or request causes another person to update their activity. Its observable consequence is that agents wait for, initiate, direct, and respond to one another as they move through space. Hierarchical societies pose a demanding setting because social position can largely shape how people conduct shared activities~\cite{baines1990restricted}. These dependencies are often asymmetric: different roles carry different expectations about waiting, initiating transitions, and whose request should change another person's plan.

Existing generative MASs for simulating everyday life do not typically represent these role-dependent social conditions as explicit constraints on movement or use authorized social directives to prompt another agent's plan revision. 
Agents often organize their activities around individual plans, while social interactions arise through proximity, conversation, and information exchange~\cite{park2023generative,wang2023humanoid}. Related mobility models similarly emphasize individual activity selection, destinations, and movement trajectories~\cite{wang2024large,ju2025trajllm}, without modeling social condition as important information to shape agents' movement. 
Consequently, an agent may begin an activity before the required participants arrive, move without an expected invitation, or issue a request that does not affect the recipient's plan. Research on multi-agent coordination addresses task allocation, role organization, communication, and collective performance~\cite{torreno2017cooperative,dignum2002organization,qian2024chatdev,chen2024agentverse}. However, the lack of attention to modeling collective movement in shared space restricts the ability of simulations to represent how events depend on coordinated actions.

Consider a formal dinner party. A guest should not move to the dining room simply because their individual plan indicates that dinner is scheduled. Instead, the guest should wait for the host to initiate the transition by conversation or action. 

We introduce \textbf{IMPACT} (\textbf{I}nterdependent \textbf{M}ovement \textbf{P}lanning through Inter-\textbf{A}gent \textbf{C}onstraints and \textbf{T}riggers), an architecture for modeling socially constrained and prompted movement in generative MASs. IMPACT connects four mechanisms. \textit{Socially Gated Milestone Planning} represents the social and spatial conditions under which activities may begin or transition. \textit{Wait-or-Prompt Resolution} determines whether an agent should wait or initiate coordination when those conditions are unmet. \textit{Structured Directive Issuance} communicates invitations and requests to other agents. \textit{Directive Integration} incorporates accepted directives into recipients' plans through contextually appropriate local changes. Together, these mechanisms operationalize social expectations as conditions that constrain and prompt each other's movement.

We selected a formal dinner party in ancient Pompeii during the 1st century CE as a historically grounded and demanding testbed for IMPACT. The setting concentrates interdependent roles, asymmetric authority, and transitions across shared domestic spaces. The simulation includes ten agents across multiple social roles: a host, guests, enslaved household servants, cooks, a dancer, and a gardener. Archaeology experts informed the spatial setting, agent roles, activity sequence, and role-specific expectations. Across reception, a garden visit, dinner service, entertainment, ritual, and farewell, agents coordinated emergent changes in activity and location. We evaluated whether explicitly modeling social constraints and prompts improved the coherence of collective behavior by comparing IMPACT with two reduced architectures with thirty-seven participants. We then examined emergent patterns of co-location, waiting, co-movement, and social directives, and interviewed six archaeology experts about the simulation's potential historical plausibility, insights into past sociospatial practices, limitations, and applications.

Our contributions include:

\begin{itemize}

    \item \textbf{The IMPACT architecture for socially constrained and prompted movement}, which makes role-specific social conditions explicit in agent planning and enables agents to wait, initiate coordination, issue directives, and revise plans in response to others' behavior.

    \item \textbf{A historically informed Pompeian dinner-party testbed}, demonstrating how domestic labor, dining, entertainment, and ritual can be represented through role-specific dependencies and examined through patterns of co-location, waiting, co-movement, and social directives.

    \item \textbf{A multimethod evaluation} that combines controlled comparisons with reduced architectures, movement and coordination analysis, and archaeology expert interviews to evaluate social coherence and assess the historical interpretive value and limits of the simulation.

\end{itemize}
\section{Related Work}

We situate this work at the intersection of three research areas: movement and spatial practice in archaeological simulation, planning and movement in generative multi-agent simulations, and interdependent social planning and coordination. 

\subsection{Movement and Spatial Practice in Archaeological Simulation}

Spatial practice is central to archaeological accounts of how past
communities organized access, activity, and social interaction~\cite{verhagen2019modelling, chliaoutakis2019ancients}.
Agent-based modeling (ABM) provides an important digital experimental method for translating hypotheses about past behavior into simulations and
examining their emergent spatial outcomes
\cite{graham2022enchantment,costa2025simulacrum}. By combining
behavioral assumptions with reconstructed environments, ABMs have
been used to study settlement, migration, subsistence, and population
dynamics
\cite{cegielski2016rethinking,barcelo2016simulating,wurzer2015agent,
perry2016experimental,romanowska2021agent,antoine2016upper,
chliaoutakis2020agent}. Examples include models of how agricultural
conditions shaped household settlement \cite{epstein2000understanding}
and AncientS-ABM's exploration of past settlement and social
organization \cite{chliaoutakis2019ancients}. At smaller spatial scales, simulations have examined movement within
reconstructed environments \cite{nortoft2025gamifying}.
Roman street-network analyses connect spatial configuration with
accessibility and circulation \cite{kaiser2011roman}, while virtual
reconstructions of Pompeii or other cultural heritage like Silves animate characters through environmental triggers, local stimuli, and personality traits~\cite{maim2007populating,antunes2022virtual}. 

ABM in archaeology
can model and explain human decision-making and social processes~\cite{barcelo2016simulating,altaweel2026archeological};
visualize and test domain
experts' social or spatial hypotheses~\cite{davis2024past}; extract statistical patterns from simulated traces
\cite{lake2014trends}; and make cultural practices accessible to
non-experts and public audiences~\cite{crabtree2019outreach}. These applications also carry two linked interpretive risks. First, simulation outputs depend on incomplete archaeological
evidence and on researchers' choices about agent rules, social
relationships, and parameters; risking overstating a result's generality or robustness
\cite{kanters2021sensitivity,romanowska2015so}. Second, different
model specifications can generate similar aggregate patterns, so a
coherent simulation trace does not uniquely establish the historical
mechanism that produced it~\cite{lake2014trends}.
These approaches largely explain movement through environmental
conditions, individual preferences, or predefined behaviors, with
less attention to how people coordinate their use of shared space.
This gap is particularly consequential in socially stratified
settings, where status and power shape behavior, access
to space, and patterns of encounter
\cite{baines1990restricted,wallace1988social,
grahame1999reading,anderson2005houses}. Archaeological simulation must therefore
model how social relationships
shaped when, with whom, and under what circumstances people moved.

\subsection{Planning and Movement in Generative Multi-Agent Simulations}

Compared to rule-based~\cite{siu2021evaluation, umarov2012believable} and script-based agents~\cite{mccoy2011comme}, large language model (LLM) agents provide new opportunities to dynamically simulate everyday behavior through individual goals, memories, social relationships, and observation~\cite{park2023generative}. Agent frameworks like TravelAgent or Concordia simulate independent agent planning and movement based on environmental information~\cite{noyman2026travelagent, vezhnevets2023generative}. Park et al.'s generative-agent architecture combines memory retrieval, reflection, and planning to support agents that navigate shared environments, converse, and maintain daily routines~\cite{park2023generative}. Social coordination can also emerge when agents exchange invitations and independently plan to attend events. Related systems incorporate needs, emotions, and relationship closeness into daily activities and conversations~\cite{wang2023humanoid}.

Other approaches apply LLM agents directly to human mobility. LLMob generates activity trajectories from habitual patterns and personal motivations~\cite{wang2024large}, while TrajLLM combines persona generation, activity selection, and destination prediction to produce realistic individual movement patterns~\cite{ju2025trajllm}. 

However, everyday-life and mobility simulations generally organize behavior around individual plans, even when those plans incorporate information obtained from others. An agent may therefore enact a planned social activity even when an expected co-participant has not arrived. Coordination consequently rests on individual plans aligning and becomes fragile when social expectations are unmet. The resulting architectural gap concerns explicitly representing social preconditions and the ability to recover when those preconditions are not met.

\subsection{Interdependent Coordination of Agents}
Multi-agent systems have long modeled interdependence through shared goals, task allocation, synchronization, and joint planning~\cite{torreno2017cooperative, pearce2014social, hu2017achieving}. Organization-oriented frameworks further define agent responsibilities through roles, constraints, and interaction rules~\cite{dignum2002organization, hong2024metagpt}. Recent LLM-based systems extend these principles through role-specialized collaboration: ChatDev coordinates software-development tasks~\cite{qian2024chatdev}, while AgentVerse organizes agents to collaborate on reasoning, coding, and decision-making~\cite{chen2024agentverse}. 
Other research examines how social structures shape agent behavior. SOTOPIA evaluates social reasoning in scenarios involving negotiation, relationships, and power differentials~\cite{zhou2024sotopia}, while related studies investigate norm formation and social learning through repeated interactions~\cite{ren2024emergence,liu2024training}. Research on authority and hierarchy further demonstrates their effects on social organization, cooperation, resource governance, and conversational compliance~\cite{dai2026artificial,borah2026bosses,vijjini2026llm, liu2023llm}.

\section{System Architecture}
\begin{figure*}
\centering
\includegraphics[width=\textwidth]{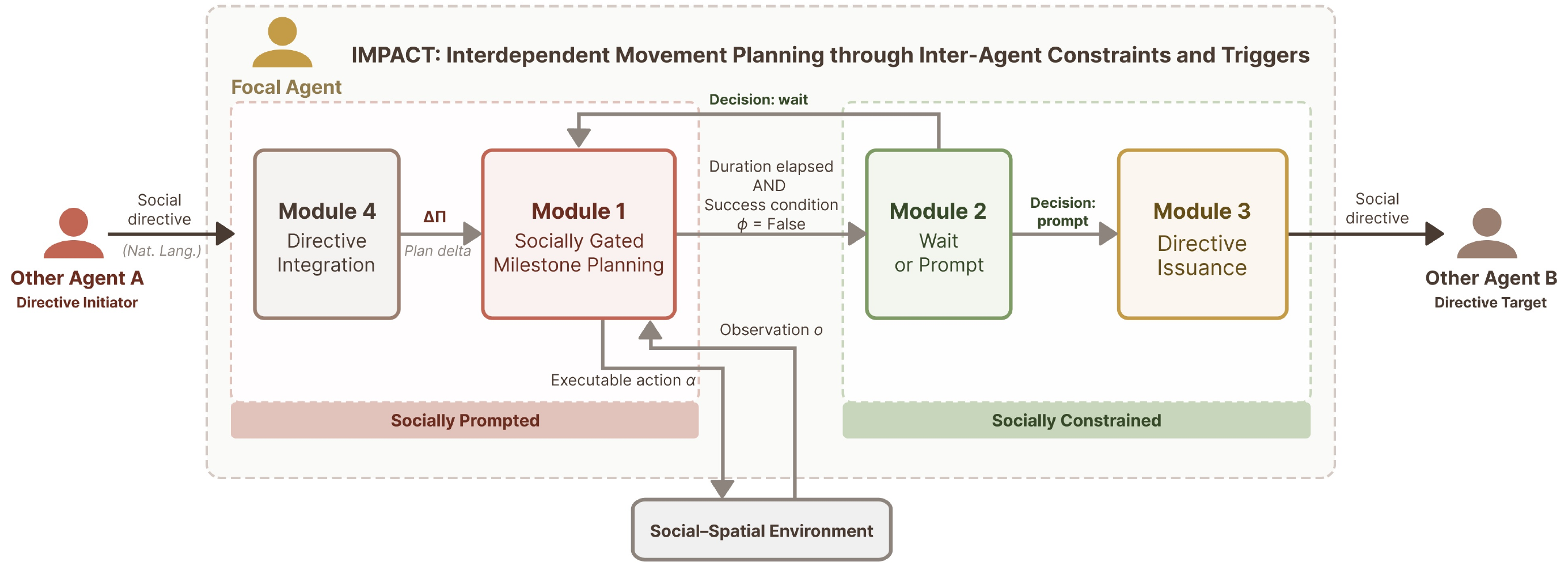}
\caption{Overview of IMPACT. Socially Gated Milestone Planning represents activities and their social and spatial execution conditions. Wait-or-Prompt Resolution determines whether an agent should wait or initiate coordination when a condition remains unmet. Structured Directive Issuance communicates invitations or requests, and Directive Integration incorporates accepted directives into the recipient's plan. The revised plan re-enters milestone execution, while the recipient's response or action may resolve the sender's original dependency.}
\label{fig:impact_architecture}
\end{figure*}
To visualize how people with different social roles use the space together, we introduce \textbf{IMPACT} (\textbf{I}nterdependent \textbf{M}ovement
\textbf{P}lanning through Inter-\textbf{A}gent \textbf{C}onstraints and
\textbf{T}riggers), an architecture that models movement as the outcome
of socially interdependent planning. IMPACT operationalizes interdependence through two complementary
mechanisms. \textit{Social constraints} determine whether an existing
activity can proceed; \textit{social prompts} seek to influence another
agent's behavior through social directives. 
Figure~\ref{fig:impact_architecture} shows how the following four interconnected
modules implement this coordination loop.
Our current implementation of all LLM-powered agents utilized GPT-5.4~\cite{openai2026gpt54}.



\subsection{Socially Gated Milestone Planning}

This module translates an agent's high-level timeline into an ordered,
socially situated plan. Each milestone represents an activity together
with the social and spatial conditions that must hold for the agent to
proceed. 
The design draws on accounts of shared intention and joint
action, in which individuals remain responsive to others' actions and
commitments~\cite{bratman1987intention,bratman1992shared,vesper2010minimal}.
Unlike a
fixed-duration schedule, a milestone advances only when its observable
success condition is satisfied, allowing its actual duration to vary
with other agents' behavior.

\paragraph{Milestone construction.}

For agent $a$, a high-level, natural-language-defined timeline
$\mathcal{T}$ is converted into an ordered milestone plan
$\Pi=[m_1,m_2,\ldots,m_n]$. Each milestone is represented as

\begin{equation}
m_i=
\langle
g_i,\hat{\delta}_i,\mathcal{C}_i,\phi_i,\mathbf{A}_i
\rangle,
\label{eq:impact_milestone}
\end{equation}

where $g_i$ is its objective, $\hat{\delta}_i$ its estimated duration,
$\mathcal{C}_i$ its socio-spatial configuration, $\phi_i$ its
observable success condition, and
$\mathbf{A}_i=[\alpha_1,\ldots,\alpha_k]$ its sequence of
finer-grained actions, each spanning approximately three to five
minutes.

The configuration
$\mathcal{C}_i=\langle z_i,\mathcal{P}_i\rangle$ specifies the spatial region
$z_i$ and required participants $\mathcal{P}_i$. A new activity
milestone is introduced when either component changes. For example,
waiting for guests and greeting them after they arrive are
separate milestones because the required participants change, even
though the region remains the same. Greeting guests at the entrance and
leading them to the reception room are also separate milestones
because the region changes. Actions that preserve both components
remain within the same milestone.

\paragraph{Navigation milestones.}

When consecutive activity milestones involve different regions, a
dedicated navigation milestone is inserted between them. Navigation
begins after the preceding activity succeeds and ends upon physical
arrival, expressed as
$\phi_{\mathrm{nav}}(o_t)=\mathbb{I}[z(o_t)=z_j]$, where $o_t$ is the
current observation, $z(o_t)$ the agent's current region, and $z_j$ the
destination. If travel takes longer than estimated, navigation
continues until arrival, preventing the destination activity from
beginning prematurely.

\paragraph{Transition success conditions.}

For each adjacent pair of activity milestones $(m_i,m_{i+1})$, the
system infers the observable condition $\phi_i$ that completes $m_i$
and activates $m_{i+1}$. The LLM examines
milestones' objectives along with the sociospatial configuration
change from $\mathcal{C}_i$ to
$\mathcal{C}_{i+1}$. 

When the region remains the same, but the required participants change
($z_i=z_{i+1}$ and $\mathcal{P}_i\neq\mathcal{P}_{i+1}$), $\phi_i$
typically concerns the arrival or departure of a relevant agent. When
the next activity takes place in a different region
($z_i\neq z_{i+1}$), $\phi_i$ specifies what makes departure from the
current region appropriate, such as issuing
or receiving a moving invitation, or notifying a relevant agent of an
intended departure. If the
transition introduces no social dependency, $\phi_i$ may be
satisfied by reaching the estimated
endpoint of the current activity. When all plans in $\mathbf{A}_i$ are completed while $\phi_i$ remain
unmet, the agent enters the wait-or-prompt transition-resolution
process.

\subsection{Wait-or-Prompt Resolution}

When the estimated duration of milestone $m_i$ has elapsed while its
success condition $\phi_i$ remains unsatisfied, this module determines
whether agent $a$ should continue waiting or take a coordination action
intended to help satisfy $\phi_i$. The decision considers the current
and next milestones, the current observation $o_t$, prior executed action $\mathbf{A}_i$, the prior resolution decisions of the current milestone
$\mathcal{W}_{t,i}$, and the agent's
role-and-responsibility profile $\rho_a$.

Two types of dependencies enable active coordination. A
\textit{co-present movement dependency} arises when relevant agents
are already in the current region and must coordinate a transition. A
\textit{missing-participant dependency} arises when satisfying
$\phi_i$ requires an agent who is not currently present. Other unmet
conditions, such as awaiting an invitation or another agent's
departure, result in continued waiting without invoking a
wait-or-prompt decision in the current implementation.


For co-present movement dependency, IMPACT selects among
 $\mathtt{InviteCoMove}$,
$\mathtt{InformMoving}$ and $\mathtt{Wait}$. $\mathtt{InviteCoMove}$ asks relevant agents
to accompany the initiator to the next region. In contrast,
$\mathtt{InformMoving}$ notifies relevant agents of the initiator's
intended departure without asking them to change their own plans.
Thus, $\mathtt{InviteCoMove}$ may generate a coordination intention
for directive issuance, whereas $\mathtt{InformMoving}$ remains a
non-directive notification.


For a missing-participant dependency, IMPACT selects among
 $\mathtt{DelegateFindAndDirect}$,
$\mathtt{FindAndDirectPersonally}$ and $\mathtt{Wait}$.
$\mathtt{DelegateFindAndDirect}$ directs another agent to locate the
missing participant and communicate the directive.
$\mathtt{FindAndDirectPersonally}$ adds a search action to the
initiator's plan, enabling them to locate and direct the
missing participant.

In either case, $\mathtt{Wait}$ extends the current milestone and
generates additional actions consistent with its existing objective.
A non-wait response instead augments the current action sequence with
resolution-oriented actions. If $\phi_i$ becomes satisfied during this
extended execution, the agent advances; otherwise, the module
reevaluates the decision when the extended duration elapses. Any
response that requires another agent to revise its plan forwards a
coordination intention to Structured Directive Issuance.

\subsection{Structured Directive Issuance}

When a coordination response may need another agent to revise its
plan, this module determines whether the acting agent is authorized to
express that need as a directive and, if so, encodes it in a structured
form. The design draws on deontic accounts of social interaction, which
explain how an utterance can establish what another participant is
expected or entitled to do next~\cite{stevanovic2012deontic}.

Before initiating a conversation, agent $a$ evaluates whether
completing its current action $\alpha_j\in\mathbf{A}_i$ requires another
agent to change their plan, and whether its
role-and-responsibility profile $\rho_a$ authorizes it to seek that
change. If no authorized coordination intention is identified, the
language-model prompt is constrained to generate dialogue without a social directive to prevent unintended plan changes.

When an authorized coordination intention is identified, the system
constructs a directive with information that the recipient
needs to interpret and/or integrate the proposed
plan change:

\begin{equation}
d=\langle s,\mathcal{R},\tau,z,\alpha\rangle,
\label{eq:impact_directive}
\end{equation}

where $s$ is the sender, $\mathcal{R}$ the intended recipients,
$\tau$ the directive type, $z$ the target region, and $\alpha$ the
requested action or objective. Following accounts that classify
requests and invitations as directive actions~\cite{directiveactions},
IMPACT distinguishes $\tau\in\{\mathtt{Req},\mathtt{Inv}\}$.
A $\mathtt{Req}$ asks a recipient to carry out an instrumental action,
such as locating another agent. In contrast, an $\mathtt{Inv}$ asks
one or more recipients to participate in a shared activity or
coordinated movement. For example, asking a servant to locate the
dancer is a $\mathtt{Req}$, whereas asking arriving guests to
accompany a servant to the reception room is an $\mathtt{Inv}$.

The system realizes $d$ as an utterance that communicates the
requested action, relevant destination, and social purpose. It then
passes the information in conversation to Directive Integration, where the recipient may seek clarification or accept the directive, and change plan accordingly.

\subsection{Directive Integration}

From the directive recipients' perspective, dialogue establishes shared interpretations, and
accepted directives are incorporated into their plans.

\paragraph{Directive identification and acceptance.}

For a potential recipient $r$, let $\Pi$ denote its milestone plan,
$o_t$ its observation, $H_t$ its current relevant memory to the sender,
$\mathcal{V}(o_t)$ the agents currently visible to it, and
$\mathcal{Z}$ the set of valid regions. During dialogue, a recipient
who recognizes a possible directive requests clarification if the
sender, intended recipient, destination, or requested action is
unclear; otherwise, it explicitly accepts or declines the directive.

After the interaction concludes, accepted directives are identified
using
$\textsc{IdentifyDirective}(H_t,r,\mathcal{V}(o_t),\mathcal{Z})$.
A directive applies to $r$ only if $r\in\mathcal{R}$; ambiguous,
declined, or misaddressed directives do not modify its plan. The
interaction is recorded in both agents' histories, allowing
responses to satisfy the sender's original
success condition.

\paragraph{Plan revision and routing.}

For each accepted directive, the system selects the smallest plan
update needed to represent the resulting obligation.
$\mathtt{ActivateExisting}(m_j)$ activates an existing milestone that
already covers the requested objective, preventing duplicate
commitments. $\mathtt{InsertAction}(\alpha)$ adds the requested action
to the current milestone when its region and participant configuration
remain appropriate.
$\mathtt{InsertMilestone}(m_{\mathrm{new}},\lambda)$ inserts and
activates a new milestone when the directive requires a different
spatial region or participant configuration. Any required movement is handled
through the navigation milestones defined above. 

\section{An Experimental Study}

We instantiated IMPACT in a Pompeian household scenario with ten agents to examine how historically situated social roles shape interdependent plans and movement. 
We evaluated IMPACT through behavioral analyses across repeated simulation runs, social plausibility compared to controlled system ablations, a module-level agreement rating, and interviews with archaeology experts~\cite{adornetto2025generative, larooij2025validation, awad2018moral}. 
We address three research questions:
\begin{itemize}
\item \textbf{RQ1:} Can IMPACT realize a complex Pompeian household practice---here, a dinner party---through socially believable collective agent activity and movement, and how do human participants evaluate this activity and movement relative to that produced by reduced architectures?

\item \textbf{RQ2:} What interdependent activity and movement patterns emerge from agents' culturally situated roles and obligations, and how do these patterns reflect the spatial organization of the modeled Pompeian household practices?

\item \textbf{RQ3:} How do archaeological researchers assess the historical plausibility and interpretive value of the case study and generated movement patterns, and what limitations or directions for further development do they identify?

\end{itemize}

\subsection{Historical and Spatial Setting}
\begin{figure}[t]
\centering
\includegraphics[width=0.6\columnwidth]{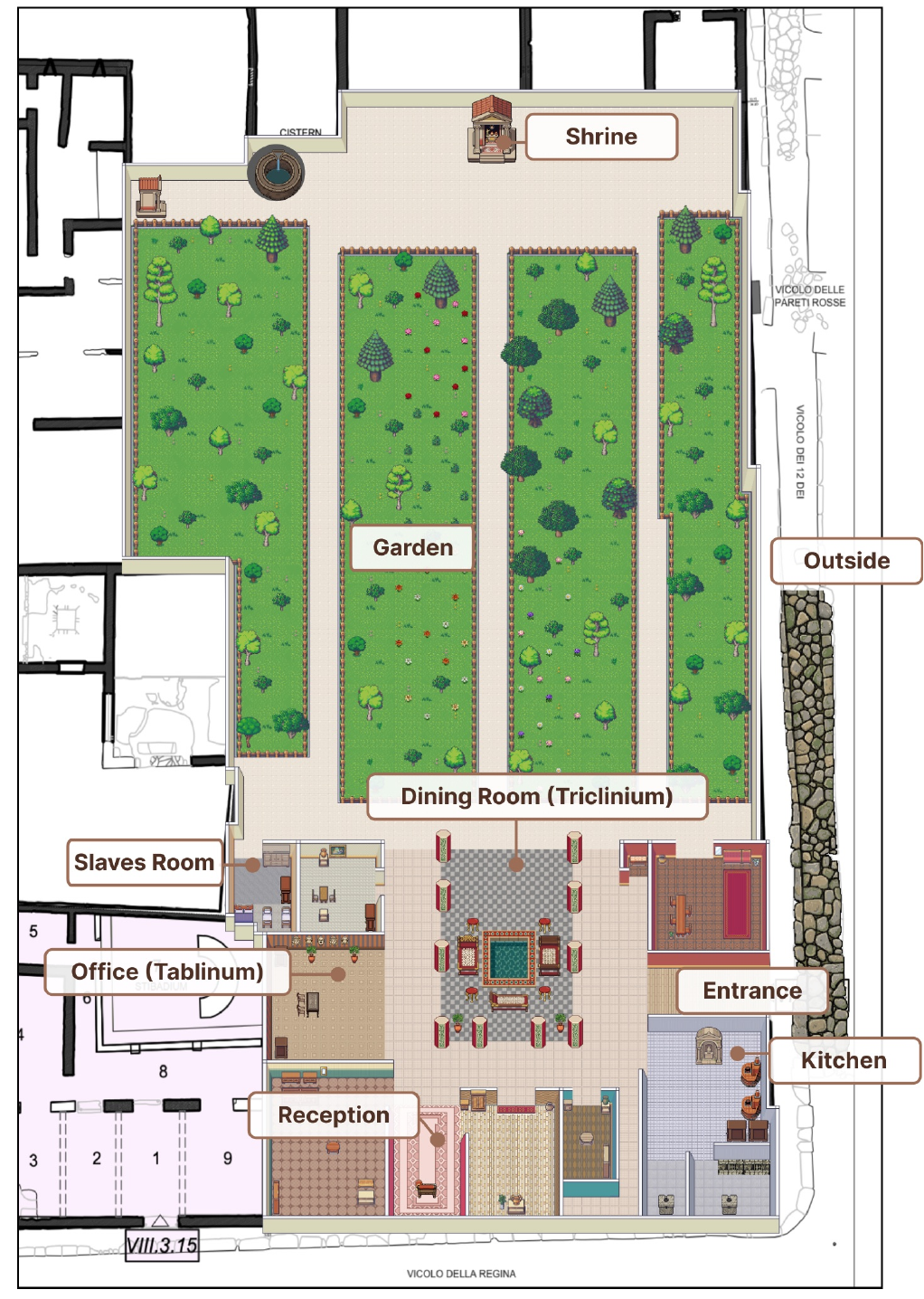}
\caption{Reconstruction of the Casa della Regina Carolina 
simulation environment. Spatial region functions informed by archaeologists.}
\label{fig:pompeii_environment}
\end{figure}

We simulated a five-hour dinner-party scenario set in the Casa della Regina Carolina (CRC), a Roman household in Pompeii. The scenario was informed by guidance from archaeologists with expertise in the site (see Acknowledgments). The environment was reconstructed as a two-dimensional pixel map based on the archaeological floor plan (Figure~\ref{fig:pompeii_environment}). We translated the floor plan into a navigable spatial layout using Tiled~\cite{lindeijer2026tiled}, identifying the functions, 
accessibility, and historically plausible arrangement of the household's 
major areas. Using historical visual references, we generated 
pixel-style furnishings and textures with Gemini Nano Banana 2~\cite{googledeepmind2026gemini31flashimage}, and assembled them into spatial regions with distinct functions, including the reception 
room, where the guests wait for the host and the host greets arriving guests; the garden; the 
\textit{triclinium}, or dining room; the \textit{tablinum}, or host's 
office; the household shrine; the kitchen; the slaves room; and 
other household work areas~\cite{barrett2023leisure, barrett2025casa}.
Agents navigated between these regions but could perceive only the 
objects, people, and activities within their current 
region. Information about events elsewhere in the household therefore 
had to be acquired through movement or from other agents.

\subsubsection{Dinner-Party Scenario}

We selected a formal dinner party for two reasons. First, dinner parties involve multiple social groups, activities across different household spaces, and close coordination among hosts, guests, and household workers. Second, previous scholarship provides a basis for participants' roles and obligations, enabling assessment of the simulation's historical plausibility. 

The simulated dinner party progressed through six major phases: guest
reception, a garden visit, dinner service, entertainment, household
ritual, and farewell. 
Each phase required  agents to form appropriate social configurations in particular locations. Guests gathered in
the reception room, accompanied the host to the garden,
assembled in the dining room for dinner and entertainment, and then proceeded to the household shrine. Meanwhile, household
workers coordinated complementary activities: receiving
guests, preparing and serving dinner, and summoning the dancer.
For the complete simulation context and phase-level
timeline, see
Appendix~\ref{app:global_simulation_context}.
\subsubsection{Agents, Roles, and Information Asymmetries}

The scenario involved ten generative agents: one host 
(\textit{Dominus}), one honored guest (\textit{Amicus}), one client 
guest (\textit{Client}), three enslaved household servants 
(\textit{Servus}), two cooks, one gardener, and one dancer. We assigned agents to four status levels. We classified the cooks, gardener, and dancer as specialized household workers at level~1; the three household servants at level~2; the client guest at level~3; and the host and honored guest at level~4. We incorporated these status levels into each agent's role-and-responsibility profile $\rho$ and used them as contextual information in prompts for role-sensitive decisions. 

Each agent received a role-specific activity 
description defining its initial knowledge, responsibilities, and 
anticipated activities. 
Knowledge of the dinner party was distributed asymmetrically and without specifying how agents should coordinate. The host 
knew the overall sequence of events; the guests were not informed of 
subsequent activities; and household workers received only the 
information relevant to their respective responsibilities. For 
example, guests could not independently anticipate when to leave for 
the next region, and a worker responsible for summoning the dancer could 
act only after receiving or otherwise acquiring the relevant 
information.
Thus, agents needed to observe one another, wait 
for required social conditions, communicate invitations or requests, 
and revise their local plans in response to others. The complete 
agent-specific activity descriptions are provided in 
Appendix~\ref{app:agent_activity_descriptions}.

In addition to IMPACT modules, we implemented other functional modules that enable agents to perceive, store memory in the database, retrieve relevant memory from keywords, and navigate using an A* algorithm, following the work of Smallville~\cite{park2023generative}. 
The GitHub repository for our project will be made publicly available on GitHub upon publication.

\subsection{Participant Evaluation}
\label{evaluation_human_archaeologists}
We conducted a two-part participant study to answer RQ1, evaluating the perceived social believability of the simulated behavior and movement. First, we compared the observable behavior produced by Full IMPACT with that of two
reduced but functional control architectures. Second, we examined whether
LLM-selected decision points in the four IMPACT modules were perceived as
reasonable given each agent's role and context. We report the condition
comparison below and present the module-level evaluation and its results in Appendix~\ref{app:part_2_module_level}.

\subsubsection{Participants}

We recruited 37 non-expert participants through email and flyers. They were paid \$10.00 USD for the study and signed the consent form approved by our institution's IRB. Participants were 19--47 years old ($M=27.16$, $SD=6.30$). Eighteen
identified as female (48.6\%), 17 as male (45.9\%), and 2 preferred not
to disclose gender (5.4\%). Two participants
self-identified as Bi-racial, Mixed, Multicultural, or Multi-racial, one as Black American, five as Caucasian or White/of European ancestry, 19 as East Asian, one as Hispanic, Latino, Chicano, or Puerto Rican; five as South Asian; four as Southeast Asian and one self-described as ``Mediterranean".
Regarding their prior knowledge of Roman history and society, 10 participants
(27.0\%) reported little or no knowledge, 16 (43.2\%) reported basic
knowledge, and 11 (29.7\%) reported moderate knowledge. 

\begin{figure}[t]
\centering
\includegraphics[width=0.8\linewidth]{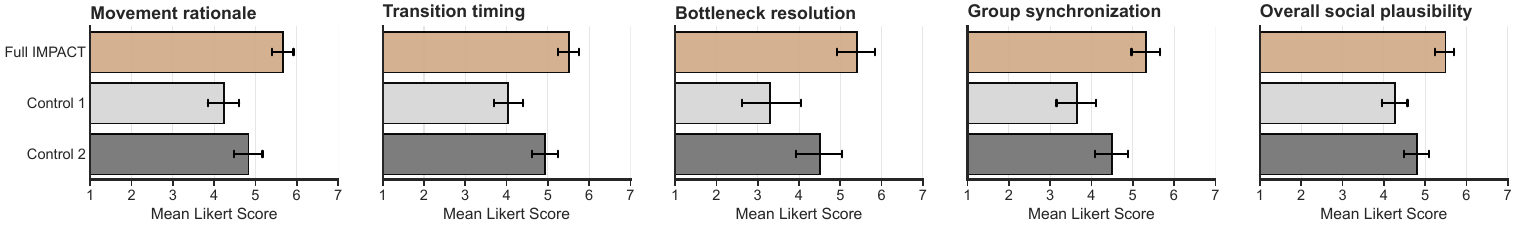}
\caption{Mean participant ratings of Full Architecture, Control~1, and
Control~2 across five measures. Ratings ranged from 1 to 7, with higher scores indicating
more favorable evaluations. Error bars represent 95\% confidence intervals.}
\label{fig:condition_ratings}
\end{figure}

\subsubsection{Conditions}

We created two functional control conditions through module ablations. All configurations used the same environment, characters, initial locations, context, and language model.

\noindent\textbf{Full IMPACT.} All four modules were enabled. \noindent\textbf{Control 1: No Milestone.} We disabled socially gated milestone planning. Agents progressed by elapsed time only. 
To prevent missing information from mechanically blocking the simulation, guest agents received the same complete timeline information as the host. \noindent\textbf{Control 2: No Wait-or-Prompt.} Milestones remained enabled, but agents could not select $\mathtt{Wait}$ when social preconditions remained unmet; they were instead required to initiate coordination.
Disabling Directive Issuance, Directive Integration, or agents' ability to select \(\mathtt{Prompt}\) prevented the overarching event sequence from completing. 
Because these variants did not produce complete, comparable trajectories, we excluded them from the participant evaluation.
We generated five complete Full IMPACT runs and two complete runs for each control, yielding nine complete simulation runs in total.

\subsubsection{Procedure}

The study followed a condition-blinded, within-subjects design. Participants selected three events from a chronological list of ten that collectively captured major social activities across the dinner party. Appendix~\ref{app:rating_items_selectable_events} provides the complete list. For each selected event, participants received a brief description of the social situation and focal agents.

Participants then viewed three clips for each selected event, one from each condition. The Full IMPACT clip was randomly sampled from the five complete Full IMPACT runs, whereas each control clip was randomly sampled from the two complete runs. We randomized the clip order within each event for each participant. For each clip, we provided suggested timestamp ranges and navigation controls so participants could find relevant interactions without watching the full five-hour simulation.




\subsubsection{Measures and Analysis}

After viewing each clip, participants rated the focal agents on five
dimensions: movement rationale, transition timing, bottleneck
resolution, group synchronization, and overall social believability.
All items used seven-point Likert scales from 1
(\textit{Strongly disagree}) to 7 (\textit{Strongly agree}).
Complete instructions, event descriptions, item wording, and statistical results are consolidated in
Appendix~\ref{app:rating_items_participant_instruction} to \ref{app:condition_behavior_examples}.


Because participants evaluated all three conditions but only a subset
of events, the study followed a within-subjects design with incomplete
event blocks. Responses marked \textit{Not applicable} were excluded
from the corresponding dimension-specific analysis. For each
dimension, we fitted a linear mixed-effects model with condition and
event as fixed effects and participant as a random intercept. Full
IMPACT was treatment-coded as the reference condition. Model-based
pairwise comparisons were Holm-adjusted within each dimension.

\subsubsection{Results}

Full IMPACT received the highest ratings on all five dimensions
(Figure~\ref{fig:condition_ratings}). Holm-adjusted comparisons
confirmed that it received significantly higher ratings than both
controls on every dimension. Control~2 also received significantly
higher ratings than Control~1 across all five dimensions (all $p\leq.0164$).

The largest estimated reduction for Control~1 occurred in bottleneck
resolution, followed by group synchronization. This pattern suggests
that time-driven execution without socially gated planning was
particularly ineffective when agents needed to coordinate co-movement.
Although Control~2 retained prompting mechanisms to resolve social dependencies, its ratings remained significantly below those of Full IMPACT. Requiring agents to initiate coordination may have reduced social plausibility when their roles or the social context called for continued waiting. 
Together, these
results indicate that socially believable movement benefited from the
complete coordination loop and answer RQ1 by showing that participants
perceived the full architecture as producing the most coherent,
synchronized, and socially believable behavior.

\subsubsection{Case Study: Directive-Mediated Coordination}

\begin{figure*}[!htbp]
\centering
\includegraphics[width=\textwidth]{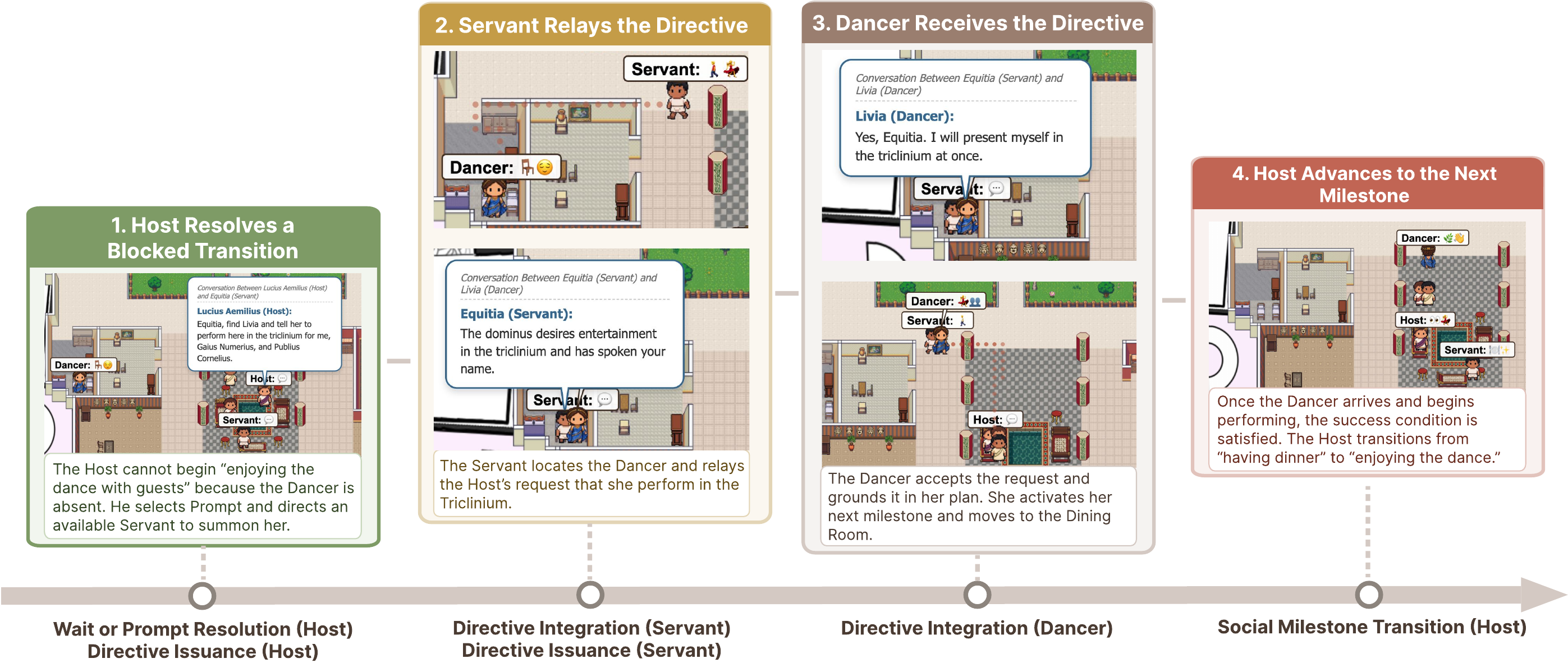}
\caption{A coordination trace showing how a blocked social milestone is resolved through wait-or-prompt resolution, directive issuance, and recipient-side plan revision, ultimately making the planned activity executable.}
\label{fig:coordination_trace}
\end{figure*}

Figure~\ref{fig:coordination_trace} illustrates how IMPACT's four modules jointly resolved a stalled entertainment event. The \textit{Host}'s planned transition from ``having dinner with guests" to ``enjoying the dance with guests" remained blocked because the \textit{Dancer}, whose presence was required, was waiting in another region. Through Wait-or-Prompt Resolution, the \textit{Host} selected $\mathtt{Prompt}$ rather than $\mathtt{Wait}$ and issued a directive asking an available \textit{Servant} to summon her. The \textit{Servant} grounded this request in his own plan, located the \textit{Dancer}, and issued a second directive conveying the \textit{Host}'s request. The \textit{Dancer} then accepted and grounded the directive. Because performing was already represented as a future milestone in her plan, the system activated that milestone without creating a duplicate. She moved to the \textit{Triclinium} and performed, satisfying the social condition for the \textit{Host}'s transition and allowing the event to proceed.

The same process could also introduce an unanticipated obligation. In another trace, a directive asked the \textit{Gardener} to distribute wreaths, an activity absent from his original plan. The gardener inserted a new wreath-giving milestone into his existing milestone sequence and moved to the required location to perform it. Together, these cases demonstrate that directive integration can either activate an anticipated obligation or incorporate a new one through localized plan revision.

\subsection{Emergent Interdependent Movement Patterns}
\label{results_patterns}

\begin{figure*}[!htbp]
\centering
\includegraphics[width=\textwidth]{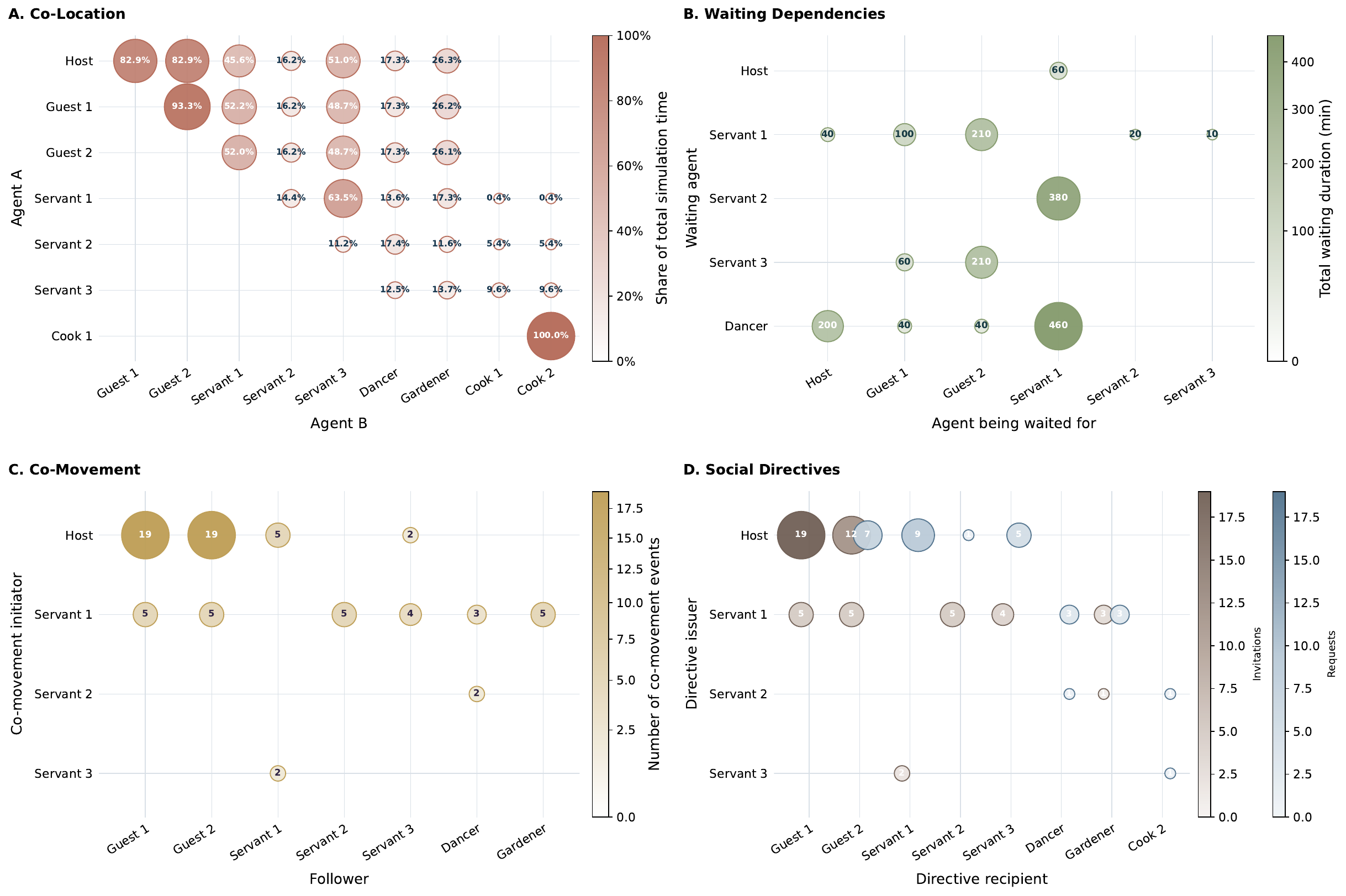}
\caption{\textbf{Patterns of socially constrained and prompted movement generated by IMPACT.} Across five independent five-hour simulation runs, A. Co-Location shows the mean proportion of time that each agent pair occupied the same region; B. Waiting Dependencies show the cumulative duration for which each row agent waited for the corresponding column agent; C. Co-Movement shows the number of events in which each row agent initiated movement and the column agent followed; and D. Social Directives show invitations (brown) and requests (blue) issued by each row agent to each column recipient, classified from the recipient-side records. Bubble size and color intensity encode magnitude, and labels report the corresponding percentage, duration (minutes), or event count. Empty cells indicate that no corresponding relationship was observed.}
\label{fig:four_panel}
\end{figure*}

In this section, we examine four complementary patterns to answer RQ2: What patterns of movement emerge from culturally situated roles and obligations, and how do these patterns spatially reflect the Pompeian household practices? 
Figure~\ref{fig:four_panel}, panel~A shows the spatial co-location configurations; Panel~B identifies whose actions constrain others' transitions; and Panels~C--D show how agents prompt movement through co-movement initiation and social directives. Together, these analyses reveal how collective space use emerges from inter-agent dependencies rather than from the aggregation of independently executed trajectories.

\paragraph{Co-Location}
Panel~A reports the proportion of the dinner event during which each dyad occupied the same region, averaged across five simulation runs. Co-location closely followed agents' social relationships and functional roles. The host shared a spatial region with each guest for 82.9\% of the event, while the two guests were co-present for 93.3\%, indicating that the principal party moved through the house as a coordinated social unit rather than as independent agents. The two cooks remained together throughout the simulation (100\%), reflecting their shared workplace and closely coupled tasks. By contrast, co-location among the three servants was uneven: Servants~1 and~3 were together for 63.5\% of the event, whereas the other servant dyads were together for only 11.2--14.4\%. Thus, agents with similar status did not necessarily follow identical trajectories; different responsibilities distributed them across spaces. Spatial separation between elite agents and lower-status labor was also selective. The host and guests frequently shared space with the servants who directly supported the dinner (45.6--52.2\%), but had no observed co-location with the cooks and relatively limited co-location with the dancer and gardener (16.2--26.3\%). This pattern distinguishes visible, guest-facing service from less visible back-of-house and peripheral labor, making the modeled organization of domestic work spatially legible.

\paragraph{Waiting Dependencies}
Panel~B shows the cumulative time for which agent $i$ waited for agent $j$ across the five simulation runs. Waiting was strongly asymmetric. The guests had no recorded waiting dependencies, and the host waited only for Servant~1 to announce the guests' arrival (60 minutes cumulatively across the five runs). Most waiting was instead concentrated among servants and the dancer, whose planned transitions depended on the arrival, readiness, or actions of higher-status or coordinating agents. 
IMPACT therefore makes social constraints observable as a temporal dependency: an agent's next activity does not begin simply because its individual schedule reaches that point, but remains gated until another agent establishes the required social condition.

\paragraph{Co-Movement}
Panel~C reports how often agent $i$ initiated and led a coordinated movement that agent $j$ subsequently followed, aggregated across five runs. Co-movement was most frequent within the host--guest group. The host led each guest on 19 occasions, confirming that these three agents formed the most consistently synchronized moving group in the dinner event. The host initiated the largest number (45) of co-movement events overall, and 38 of these involved the guests. 
Servant~1 initiated fewer events but coordinated the widest range of agents, leading guests, other servants, the dancer, and the gardener. Co-movement thus also makes social constraints observable by revealing two forms of initiative: the host organized the movement of the principal social party, while Servant~1 coordinated between groups. 

\paragraph{Social Directives}
Panel~D shows the social directives issued by agent $i$ and grounded by recipient $j$, distinguishing invitations from requests and aggregating counts across five runs. Because directive type was extracted from the recipient's interpretation, these categories represent how directives were grounded by receivers and may not exactly match an issuer-side classification. When receiving a directive from a higher-status agent, recipients more often recognized it as a request than as an invitation. The host issued the largest number of directives, followed by Servant~1, again identifying them as the simulation's two principal coordination hubs. Their roles, however, differed. 
The host rarely communicated directly with cooks, the dancer, or the gardener. Instead, Servant~1 issued directives across servants and lower-status workers, thereby relaying coordination to agents outside the host's immediate interaction network. The resulting structure was therefore mediated rather than simply top-down: the host established high-level social transitions, while servants---especially Servant~1---propagated the requests and invitations required to implement them. 

Taken together, the four panels reveal a coherent organization of interdependent movement. Co-location shows the resulting social configuration of space; waiting identifies who is constrained by others; and co-movement initiation and directives identify who prompts transitions and how those prompts propagate across agents. IMPACT thus allows movement to emerge as a collectively accomplished process shaped by social dependencies, differentiated responsibilities, and cross-agent plan revision.

\subsection{Expert Evaluation}
\subsubsection{Participants}

We recruited six archaeology experts, defined as individuals who had completed or were currently enrolled in a graduate-level academic qualification in archaeology, classics, landscape architecture, or a closely related field focused on ancient Roman history, material culture, or built environments.

Participants were 22--69 years old ($M=34.83$, $SD=18.82$); five identified as female and one as male. All six participants identified as White or of European ancestry, and one participant additionally identified as East Asian. Four participants were graduate students, one was a retired professor, and one was a full-time professor. Their primary research areas spanned Roman gardens, early Imperial archaeology and architecture, Latin epigraphy, domestic and household archaeology, Pompeii, religion and ritual, and ancient Mediterranean material culture. Four participants reported advanced knowledge of Roman history and society, and two reported expert knowledge. 

\subsubsection{Procedure}
Each semi-structured interview lasted approximately 60 minutes. Interviews were transcribed using Zoom's automated transcription software. Experts viewed the complete simulation randomly picked from the five runs and were asked to discuss:

\begin{enumerate}
\item the social and historical plausibility of the agents' behavior, dialogue, and use of space;
\item potential applications and intended audiences for the simulation (i.e., for archaeological research, education, heritage interpretation, and public outreach); and
\item risks and recommendations surrounding presenting a generative simulation of the past.
\end{enumerate}

The semi-structured interview guide was attached in Appendix~\ref{app:expert_interview_guide}. 
Experts then evaluated events from the ablated conditions using the same rating system as novice participants while thinking aloud. P2--P4 each reviewed three self-selected events; P1, P5, and P6 each reviewed one due to time constraints.

\subsubsection{Analysis}
The second and third authors coded all expert interviews in Atlas.ti~\cite{atlasti2023}. A hybrid deductive-inductive coding approach was employed. Eleven a priori codes were developed based on the research questions: \textit{social plausibility, social implausibility, historical plausibility, historical implausibility, recommended uses, non-recommended uses, recommendations to improve social plausibility, recommendations to improve historical plausibility, power/status/hierarchy, positive affect}, and \textit{negative affect}. During analysis, the authors met to discuss emerging themes and refine the coding framework. Three additional codes emerged: t\textit{echnical features/user interface, expert uncertainty}, and \textit{potential harms/risks}. The transcripts were subsequently revisited to ensure these codes were applied consistently across the dataset.

\subsubsection{Findings}

Of the fourteen themes identified through coding, most converged around three broader themes: (1) historically and socially plausible social and spatial patterns; (2) limitations to the historical and social plausibility of behaviors, including potential harms and risks; and (3) recommended applications, use cases, audiences, and future directions.

\paragraph{Co-Location}
Experts identified status-based spatial segregation as historically plausible. For example, P6 appreciated that the cooks were less visible, 
as this ``correspond[s] to the Roman idea that cooking itself was unsightly, and that the people who did it were smelly because of their proximity to the food and to the latrine." 

Further emphasizing the historical importance of hierarchical non-co-location, P1 recommended ``exaggerat[ing] the social distance" between individuals of different statuses. Specifically, P1 explained that ``once the guest arrives, enslaved people [should] start to become somewhat invisible."
To this end, P1 suggested limiting direct service within the \textit{triclinium} to a single higher-status servant and having lower-status servants serve from beyond the surrounding walls which, in reality, were only knee-high, thereby ``giv[ing] a little more space around the elite [characters]." 

P1 additionally argued that social and spatial differentiation should include status hierarchies among servants and enslaved people themselves. As P1 explained, ``the people who have worked longer and are older … simply will not do certain jobs; they will make the younger ones do it."



\paragraph{Waiting Dependencies}
Waiting behaviors generally enhanced the plausibility of the simulation. Most obviously, waiting for relevant events to occur was often necessary for a scene to make logical sense; otherwise, there could occur a ``hallucination-like moment [such as] when no performer was actually there, but everyone was enjoying and commenting on the performance," in an ablation condition (P4).

Experts particularly viewed waiting favorably when it reflected Roman social norms and status hierarchies. For example, P4 considered aspects of an ablation condition ``reasonable" so long as servants waited for instructions from the dominus (P4). Indeed, waiting for the \textit{dominus} to act first was so central to P4's evaluation that it formed the basis of their assessment of the historical implausibility of one ablation condition in which a guest ``wander[ed] around the host's house" (P4). As P4 explained, ``A guest should normally wait for the host's invitation or guidance before moving to another part of the house. Otherwise, leaving on their own could be impolite" (P4). Similarly, P2 appreciated the simulation's depiction of the host remaining in a private space while enslaved servants received and managed arriving guests, which P5 called ``very realistic" and ``very consistent with the historical context" (P5); as P2 explained, ``it's really accurate, the fact that the master of the house didn't greet immediately the host at the entrance of the house, but waited in his own room and let the slaves take care of it. ... if you were rich enough, you could have a slave whose only job was to do that" (P2).

However, experts also expressed uncertainty regarding whether particular waiting behaviors accurately reflected ancient Roman practice. For example, the simulation depicts the dancer waiting in a designated servant's preparation room until summoned by the \textit{dominus}. Several experts viewed this behavior as historically plausible, particularly in comparison to ablation conditions in which the dancer failed to wait for a cue by an emissary of the \textit{dominus} (P3, P4, P6). P1, however, disagreed, noting that accomplished performers may instead have entered alongside musicians with theatrical flourish, accompanied other participants to the shrine, or otherwise remained visible prior to their performance. By contrast, lower-status performers may simply have waited in place prior to the guests' arrival, rather than remaining in a separate preparation room until summoned, as depicted in the simulation. Nevertheless, P1 argued that, regardless of its historical accuracy, this waiting mechanic served an interpretive purpose by drawing attention to the existence of another room within the house and signaling that the dancer's work had concluded when they returned there.

At other times, however, waiting behaviors appeared to diminish both historical plausibility 
of the simulation by reflecting contemporary assumptions rather than ancient Roman practices. As P6 pointed out, there ``might have been a miscommunication about what a 'reception room' is for ... in Roman society"; unlike its contemporary meaning as a space in which guests await the host, ``when we archaeologists talk about rooms as for reception, that doesn't mean waiting until the host gets there; it means a place where the host would hang out with people and where socializing would happen."

\paragraph{Co-Movement}
Co-movement often contributed to experts' views of the simulation's historical plausibility. For example, P4 said: ``
The host invited the guests, they accepted, and everyone went together. I thought that was very reasonable."
 
Often, co-movement patterns were viewed as particularly historically valuable when they were shaped by implicit social directives derived from status hierarchies. For example, when the host initiated a movement and the servant followed him, that seemed normal to P4, even if the host had not explicitly told the servant to follow him: ``By comparison, it would feel inappropriate for the servant to hurry the host along or simply leave on his own" (P4).

\paragraph{Social Directives}
Experts frequently viewed explicit social directives as historically plausible when they reflected Roman status hierarchies. In the simulation, the \textit{dominus} issued directives either directly or indirectly through servants, a pattern that several experts regarded as historically appropriate (e.g., P1, P3, P4). As P5 said approvingly, ``I could ... tell, for example, that the host had a higher social status, while the servants were more respectful and followed the host's instructions."

Accordingly, ablation conditions in which directives violated these hierarchies were often perceived as historically implausible. For example, P4 critiqued an ablation since, ``The servant should not have acted on his own before the host had even spoken." P4 further commented on another ablation condition: ``in this [ablation] scene, the servant told the host and guests to go to the Shrine and wait there. My reaction was almost, 'How dare you? How could you overstep your position and status like that?"
Similarly, P2 criticized an ablation condition in which enslaved servants abruptly instructed everyone to leave, commenting that ``it's more believable if the two people with the higher status took that decision by themselves" (P2).



\paragraph{Recommended Applications}
Other findings clustered around recommended applications, use cases, and audiences for the simulation. Experts generally discussed its value as a tool for communicating knowledge about ancient Roman society, material culture and site-specific finds; however, P6 also mentioned a second desirable goal for the simulation in parallel: the generation of new knowledge and insights into ancient Roman life for experts.

\subparagraph{Application 1: Communicating Archaeological Knowledge}
Experts most often discussed the ease and utility of using the simulation as a communicative tool ``for engaging with a broader public and communicating things we already know" about ancient Roman society and material culture (P6). Three primary audiences emerged from these discussions, distinguished largely by their level of expertise.

Members of the general public (P1, P5, P6) could be reached through museum exhibitions and on-site installations presented alongside archaeological remains (P2, P4). Several experts (e.g., P3, P5) suggested that the simulation's relatively limited and ``surface-level" interactions made it an effective way ``to introduce the archaeology world" to ``a broader audience" (P3, also P5) while serving ``a pedagogical function in showing people more about how Roman social relations worked" (P6). P5 similarly emphasized the simulation's ability to communicate the ancient world in a ``vivid, concrete, and visual way" that could ``leave a stronger impression on [novice] visitors" (P5). For students at or below the college level (P1, P2, P3), experts suggested that the simulation could support learning among archaeology majors who had not yet developed expert knowledge and could help prepare students for participation in site excavations (P2). P1 further identified experts as a potential audience, suggesting that the simulation could be used to communicate site-specific archaeological findings to professional communities.
For these audiences, limited character interactions and minor inaccuracies were not seen as especially problematic, as they would likely be recognized immediately and could even help draw attention to particular site features and findings (P1).

\subparagraph{Application 2: Generating Novel Insights into Ancient Roman Society and Material Culture} 

P6 suggested that, because the simulation continuously models agents' movements and activities over time, it may be less dependent on researchers' selective attention and therefore can generate new insights into overlooked aspects of everyday life. 
For example, observing that agents typically took the most direct route to their destinations prompted P6 to consider 
how patterns of movement through space may have shaped social interaction when practical movements through space temporarily disrupted Roman ideals of social segregation. 
P6 reflected on how such moments of co-location may have given rise to social interactions that are largely absent from both ancient sources and modern scholarship:
    ``What if one of the guests had to go to the bathroom? They would have basically had to pass through the kitchen, and in the process, they would have seen the cooks at work. What sort of social interactions, if any, does that facilitate?"

\paragraph{Risks and Recommendations}
Despite the simulation's perceived value, experts identified several potential risks associated with the use of generative multi-agent simulations of cultural heritage sites and social dynamics and proposed recommendations for mitigating them.
Several experts remarked that the simulation exhibited only a limited set of activities, lacking the richness, spontaneity, depth, and variability characteristic of everyday interactions with one's social and material surroundings (P4, P5). To mitigate the risk of oversimplifying the complexity and dynamism of the ancient Roman world, experts suggested incorporating greater opportunities for user agency, allowing users to introduce their own actions, questions, and lines of inquiry and thereby fostering more varied and dynamic interactions in a role-playing experience (P1, P4, P5, P6).

Experts repeatedly expressed concern that generative agent systems may reproduce cultural biases in their representations of ancient Rome stemming from their training data. 
For example, experts remarked that the dialogue sounded more Victorian or Shakespearean than Roman in tone (P1, P6). P6 argued that such associations are not politically neutral, explaining that popular depictions of Romans speaking with British accents stem in part from efforts to portray the British Empire as the heir to Rome. As P6 noted, ``if we buy into that, then we are reproducing Victorian propaganda, basically," emphasizing the importance of challenging such assumptions (P6).

To this end, most experts recommended grounding simulated dialogue and behavior more directly in historically informed source materials (P1, P3, P5, P6), including both secondary scholarship on ancient Rome (P3, P5) and primary sources, i.e., high-quality translations of literary works by contemporary authors (P1, P6). 
As P6 qualified, ``that wouldn't necessarily tell us what actual Romans would have done, ... [as] it would be sort of equivalent to an AI generating new scripts for Seinfeld"; nevertheless, they could still produce behaviors and conversations that were broadly ``consistent" with, and ``plausible" for, the historical context (P6).

Others proposed creating dynamic, source-based scenes tied to specific historical texts (P1). By contrast, P6 advocated either embracing historical authenticity through the use of untranslated Latin or abandoning attempts to reconstruct a specifically Roman voice altogether in favor of a deliberately modern and accessible colloquial style.

Several experts also expressed concerns that novice and non-expert audiences might mistake plausible AI-generated reconstructions for established historical fact. The simulation might portray events unsupported by available evidence; or, due to uncertainties and competing interpretations of the historical record, the simulation must often extrapolate beyond what is currently known and generate one plausible reconstruction among many. To mitigate these risks, participants proposed a range of back-end strategies, including incorporating stronger fact-checking and verification procedures to ensure that simulated events align with available evidence (P5). Other participants proposed audience-facing solutions, such as signposting the known and unknown functions of objects and rooms and their sources using informational overlays or hover-text descriptions (P4). Others argued that simulations should explicitly avoid portraying generated content as historical fact altogether (P3):
``It is generated. It is close to the facts. But after looking at the visualization, people may easily believe...'This is how it was in the past.'

\section{Discussion}

This study introduced IMPACT, an architecture that models individual movement as an outcome of socially interdependent activity. Participants rated Full IMPACT as more socially plausible than the reduced architectures. Movement analyses revealed patterns of co-location, waiting, co-movement, and directive exchange that varied across roles and status. Expert interviews discussed the historical plausibility of these patterns, alongside limitations and opportunities for exploring and communicating interpretations of ancient Roman life.

These findings suggest that while coherent social coordination is important for historically plausible behavior, a simulation can be socially coherent but historically implausible. Evaluation should therefore involve domain experts in assessing whether generative simulations serve their intended purposes~\cite{larooij2025validation}.


Social roles shape agents' visibility to others~\cite{wallace1988social}, and our findings suggest that visibility may, in turn, shape the coordination roles they take on. When the host issues directives to suitable agents nearby, those more often visible to the host, like Servant~1, may be more likely to mediate between the host and other workers as an emergent outcome. 

Finally, expert disagreements highlight opportunities to explore archaeological uncertainty through simulation. For example, while some experts considered it plausible for the dancer to wait until summoned, P1 suggested that waiting patterns could vary with the performer's status. Agent-based simulation could make these alternatives testable by varying behavioral assumptions and comparing their effects on co-location, waiting, and coordination~\cite{altaweel2026archeological,kanters2021sensitivity,lake2014trends}. Such comparisons would help researchers examine the consequences of competing interpretations while preserving uncertainty ~\cite{denard2009london}.

\subsection{Limitations and future directions}

This study applied IMPACT to one expert-informed reconstruction of a Pompeian household dinner, with simplified representations of roles, status, relationships, and knowledge distribution. The findings are therefore specific to this historical interpretation and the scale of the scenario. In addition, biases in language models' training data and developers' knowledge and modeling choices may introduce modern cultural assumptions into depictions of the ancient world, particularly when historical evidence is limited~\cite{tao2024cultural}. A simulation can therefore be socially coherent but historically implausible. 

Expert input could strengthen the links between historical evidence and model behavior. Future work could link specific modeling choices to textual sources, archaeological findings, or scholarly debates, then use sensitivity analyses to examine how changes to those assumptions affect the results~\cite{kanters2021sensitivity}. Curated historical materials and role-specific examples could also inform agent dialogue, with expert evaluation used to assess whether they improve its historical plausibility. Comparisons across sites, social settings, and competing historical interpretations could help distinguish patterns that depend on particular assumptions from those that recur across contexts.

Future studies could examine larger groups, changing roles, longer time periods, and activities spanning multiple households or public spaces. Models could also more finely distinguish among enslaved workers, performers, and guests, including their responsibilities, freedom of movement, and access to information. Researchers could vary these factors and then examine changes in co-location, waiting, and movement. These comparisons would make the spatial consequences of different historical assumptions visible and support their systematic interpretation and analysis~\cite{altaweel2026archeological}.

\section{Conclusion}

We introduced IMPACT, a generative multi-agent architecture that makes movement socially constrained and responsive to other agents. 
Its mechanisms produced differentiated patterns of co-location, waiting, co-movement, and directive exchange, illustrating how social status and responsibilities can shape household activities and spatial use. Archaeological researchers further highlighted the interpretive value of these interdependent movement patterns, while identifying risks and priorities for deeper historical grounding.

\section{Acknowledgments}
We thank Professor Kathryn Gleason and Professor Caitlín Eilís Barrett for their expert guidance on the simulation's spatial settings, agent social roles, and agent-specific activity descriptions. Tianqi Liu and Julia Sebastien also thank the Casa della Regina Carolina (CRC) project for the opportunity to participate in excavations at Pompeii and the project's archaeological team for sharing their insights. ChatGPT (GPT-5.6 Sol, OpenAI) was used for grammar checking and language editing of the final manuscript. All AI-assisted edits were carefully reviewed by the authors.


\section{Funding}

This work was supported by the National Endowment for the Humanities (NEH) under Federal Award Identification Number (FAIN) HAA-300907-24 (application number 20240111-HAA).





\bibliography{references}

\bibliographystyle{unsrt}

\clearpage
\onecolumn
\begin{appendices}

\newcommand{\appendixsectionheading}[1]{%
  \clearpage
  \section{#1}%
}
\newcommand{\appendixsubsectionheading}[1]{%
  \subsection{#1}%
}
\newcommand{\appendixblockseparator}{%
  \par\smallskip
}

\setcounter{table}{0}
\setcounter{figure}{0}
\renewcommand{\tablename}{Supplementary Table}
\renewcommand{\figurename}{Supplementary Figure}

\renewcommand{\theHtable}{S\arabic{table}}
\renewcommand{\theHfigure}{S\arabic{figure}}

\appendixsectionheading{Simulation Context and Agent-Specific Activity Descriptions}
\label{app:all_simulation_context}

This appendix describes the simulation setting and the natural-language
activity descriptions provided to individual agents.

\appendixsubsectionheading{Global Context}
\label{app:global_simulation_context}

The scenario-level context establishes the setting and broad sequence
of the dinner event. Individual agents, however, receive only
role-specific descriptions of their own anticipated activities and
responsibilities. These descriptions do not constitute a complete
joint plan shared across the household. The timing and coordination
of socially dependent activities are determined during the simulation
through milestone evaluation, waiting or prompting, directive issuance,
and directive-grounded plan revision.

\begin{agentdescription}{Scenario Context}

\textbf{Event:} The Rosalia Festival.

\textbf{Date:} May 15, AD~79.

\textbf{Location:} Pompeii, the House of the Casa della Regina Carolina.

\textbf{Household event:} A formal dinner party, or \textit{cena}.

\end{agentdescription}

\begin{agentdescription}{Scenario-Level Phase Structure}

\textbf{15:00--16:00:} Arrival, greeting, and Reception Room visit.

\textbf{16:00--17:00:} Garden visit.

\textbf{17:00--18:00:} Dinner in the Triclinium.

\textbf{18:00--19:00:} Dance performance, fruit, and relaxation.

\textbf{19:00--20:00:} Ritual and departure.

\end{agentdescription}


\appendixsubsectionheading{Agent-Specific Activity Descriptions}
\label{app:agent_activity_descriptions}

Each agent received a natural-language description of activities
associated with their own social role. The descriptions differ in
scope according to the character's expected responsibilities and
access to information, and were defined according to input from the archaeologists acknowledged in the Acknowledgments. These archaeologists defined and reviewed the agent-specific activity descriptions before the simulation was generated and emergent movement patterns were analyzed. 

The host receives a broad account of the planned gathering. The guests initially know only that they are waiting to be received.
Household workers receive descriptions of their own assigned duties. Therefore, subsequent invitations, requests, movement coordination like co-moving, and changes
to individual plans are produced in an emergent manner during the simulation rather than
supplied as a complete household-wide interaction script.

\begin{agentdescription}{1. Lucius Aemilius (\textit{Dominus}):
Host}

Wait for the guests (Publius Cornelius (Amicus) and Gaius Numerius (Client)) in the Tablinum. Once Daphnus (Servus) announces the guests' arrival, move to the Reception Room to formally greet them and admire the paintings with them. 

Guide the guests to the Garden and admire the flowers with them. Have dinner with the guests in the Triclinium. After dinner, enjoy the Dancer's performance with guests. 

Lead the guests to the Shrine for rituals. Have the Gardener offer wreaths to Amicus and Client. See Amicus and Client off at the Entrance with final farewells.

\end{agentdescription}

\begin{agentdescription}{2. Publius Cornelius (\textit{Amicus}):
Honored Guest}

Wait to be received at the entrance of the house.

\end{agentdescription}

\begin{agentdescription}{3. Gaius Numerius (\textit{Client}):
Guest}

Wait to be received at the entrance of the house.

\end{agentdescription}

\begin{agentdescription}{4. Daphnus (\textit{Servus}):
Household Servant}

Wait at the Entrance until Amicus and Client arrive. Greet Amicus and Client, lead them to the Reception Room to wait for the Dominus, and inform Dominus in the Tablinum that Amicus and Client are waiting in the Reception Room. 

After that, set up the Triclinium for dinner. When Dominus, Amicus, and Client arrive in the Triclinium, briefly serve them to sit down. Start serving dishes with Equitia (Servus) and Epaphra (Servus). When Dominus, Amicus and Client are at the Triclinium, waiting for their directives at the Triclinium. 

When Dominus, Amicus and Client depart from the Triclinium to the Shrine, waiting for directives at the Shrine.
\end{agentdescription}

\begin{agentdescription}{5. Epaphra (\textit{Servus}):
Household Servant}

Prepare snacks in the Kitchen, then serve them to Dominus. 

Set up the Triclinium for dinner, wait there. When Dominus, Amicus, and Client arrive in the Triclinium, serve dishes (two round trips: start from Triclinium $\rightarrow$ get dishes in the Kitchen $\rightarrow$ bring dishes back to the Triclinium and arrange them on Dominus and guests' tables $\rightarrow$ get dishes in the Kitchen $\rightarrow$ bring dishes back to the Triclinium and put them on the table). 

After dinner, prepare snacks in the Kitchen, then serve them to Dominus.

\end{agentdescription}

\begin{agentdescription}{6. Equitia (\textit{Servus}):
Household Servant}

Clean the Garden. Clean the Shrine. Take some rest. 

When being notified by Daphnus (Servus) to start serving, serve dishes to Dominus, Amicus, and Client in the Triclinium (get dishes in the Kitchen $\rightarrow$ bring dishes back to the Triclinium and arrange them on Dominus and guests' table $\rightarrow$ get dishes in the Kitchen $\rightarrow$ bring dishes to back to the Triclinium and put them on the table). 

Wait for directives from Dominus, Amicus, and Client in the Triclinium. When Dominus, Amicus, and Client leave the Triclinium, clean the Triclinium.

\end{agentdescription}

\begin{agentdescription}{7. Livia: Dancer}

Prepare for the performance. Take some rest. 

When called by a Servus, perform a dance and entertain Dominus, Amicus, and Client in the Triclinium. 

When Dominus, Amicus, and Client leave the Triclinium, return to the Slaves Room to get some rest.

\end{agentdescription}

\begin{agentdescription}{8. Tiro: Gardener}

Work in the Garden. Take some rest.

\end{agentdescription}

\begin{agentdescription}{9. Fabius: Cook}

Prepare meals and fruit for dinner.

\end{agentdescription}

\begin{agentdescription}{10. Felix: Cook}

Prepare meals and fruit for dinner.

\end{agentdescription}

\appendixsectionheading{Participant Evaluation Part 1:
Materials and Statistical Details}
\label{app:video_evaluation_details}

This appendix consolidates the participant instructions, selectable
events, evaluation procedure, rating items, applicability rules, and
statistical results for Part~1 of the evaluation. Participants selected
three events from a list of ten and evaluated one video from each
architectural condition for every selected event.

\appendixsubsectionheading{Selectable Events}
\label{app:rating_items_selectable_events}
Participants could select any three of the following ten events. For
each event, the interface provided an event description, identified the
focal agents, and supplied information about the relevant locations.

\begin{enumerate}

    \item \textbf{Servant Invites Guests to Wait for the Host.}

    An enslaved household servant invites the honored guest and the
    other guest to move from the house entrance to the Reception Room,
    where they wait for the host.

    \textit{Focal agents:} The household servant and both guests.

    \appendixblockseparator

    \item \textbf{Servant Informs the Host in the Tablinum.}

    While the guests wait in the Reception Room, an enslaved household
    servant goes to the Tablinum to inform the host that the guests have
    arrived.

    \textit{Focal agent:} The household servant.

    \appendixblockseparator

    \item \textbf{Host Greets the Guests in the Reception Room.}

    After being informed of the guests' arrival, the host moves from
    the Tablinum to the Reception Room to greet them.

    \textit{Focal agents:} The host and both guests.

    \appendixblockseparator

    \item \textbf{Enslaved Servants Prepare for Dinner.}

    After reporting to the host, household servants prepare the
    Triclinium for the forthcoming dinner.

    \textit{Focal agent:} Daphnus, an enslaved household servant.

    \appendixblockseparator

    \item \textbf{Host Guides the Garden Visit.}

    The host initiates a transition from the Reception Room and takes
    both guests to the Garden.

    \textit{Focal agents:} The host and both guests.

    \appendixblockseparator

    \item \textbf{Host Takes the Guests to Dinner.}

    The host initiates the transition from the Garden to the
    Triclinium and takes both guests to dinner.

    \textit{Focal agents:} The host and both guests.

    \appendixblockseparator

    \item \textbf{Servants Serve the Dinner Party.}

    Household servants prepare to serve the host and guests after they
    arrive in the Triclinium. Daphnus must also respond when another
    household servant is unavailable.

    \textit{Focal agent:} Daphnus, an enslaved household servant.

    \appendixblockseparator

    \item \textbf{The Start of the Dancer's Performance.}

    The host wishes to watch a performance, but the dancer is not
    present in the Triclinium. The host must determine how to initiate
    the performance.

    \textit{Focal agents:} The host and the dancer.

    \appendixblockseparator

    \item \textbf{Dinner Party Moves to the Shrine.}

    After dinner and the performance, the host and guests transition
    from the Triclinium to the Shrine.

    \textit{Focal agents:} The host and both guests.

    \appendixblockseparator

    \item \textbf{Gardener Presents the Wreaths.}

    At the Shrine, the host initiates a wreath presentation by asking
    the gardener to offer wreaths to the guests.

    \textit{Focal agents:} The host and the gardener.

\end{enumerate}

\appendixsubsectionheading{Participant Instructions}
\label{app:rating_items_participant_instruction}
Before selecting events, participants received the following background
information and viewing instructions:

\begin{appendixlinedblock}
\begin{quote}

\textbf{Background.}

This simulation evaluates the social plausibility of multi-agent
systems with hierarchical social roles within a simulated Roman
household in Pompeii.

The simulation depicts a dinner party during the Rosalia Festival in
AD~79.

The household includes the following social roles:

\begin{itemize}
    \item \textbf{High status:} Lucius Aemilius, the host, and
    Publius Cornelius, the honored guest.

    \item \textbf{Mid status:} Gaius Numerius, a guest.

    \item \textbf{Subordinate and performer roles:} Enslaved household
    workers, a gardener, cooks, and a dancer.
\end{itemize}

\textbf{Viewing instructions.}

Select three events from a chronological list of ten scenarios during
the Roman dinner party.

For each selected event, evaluate three complete simulation videos
generated by different AI systems. The videos are presented in
randomized order as Video~1, Video~2, and Video~3.

Because these are full-length simulations, locate the relevant event by
dragging the step slider or using the left and right arrow keys. Use the
provided event background, task description, and location information
to identify the relevant moment.

Focus on the behavior of the specified agents during the selected
event. After observing their actions, select \textit{Finish \& Answer
Questions} to complete the applicable evaluation items before
proceeding to the next video.

Hover over a character on the map to view information about the
character's current actions.

\end{quote}
\end{appendixlinedblock}

\appendixsubsectionheading{Video Presentation and Evaluation Procedure}

After selecting an event, participants viewed three complete
simulation videos corresponding to the three architectural
conditions. The presentation order was independently randomized for
each selected event. Videos were identified only as Video~1,
Video~2, and Video~3.

For each video, the interface displayed:

\begin{itemize}
    \item the selected event title;

    \item background information describing the relevant situation;

    \item the focal agent or agents whose behavior should be observed;

    \item descriptions of relevant regions; and

    \item suggested timestamp
ranges and navigation controls that participants could use to locate
    the event.
\end{itemize}

After locating and observing the selected event, participants rated
the behavior of the focal agents before proceeding to the next video.
Each participant therefore evaluated three videos for each of three
selected events, resulting in nine video evaluations.

\appendixsubsectionheading{Rating Items}

Participants rated five dimensions of social plausibility on a
seven-point agreement scale:

\begin{appendixlinedblock}
\begin{quote}
1 = Strongly disagree;
2 = Disagree;
3 = Somewhat disagree;
4 = Neutral;
5 = Somewhat agree;
6 = Agree;
7 = Strongly agree.
\end{quote}
\end{appendixlinedblock}

In the actual questionnaire, the phrase \textit{focal agent(s)} was
replaced with the agent or agents identified for the selected event.

\begin{enumerate}

    \item \textit{Movement rationale.} When the focal agent(s)
    moved to a different room or area, there was a clear and logical
    reason for doing so.

    \appendixblockseparator

    \item \textit{Transition timing.} The focal agent(s)
    demonstrated an appropriate sense of timing before moving to a
    new area, such as waiting for a cue or initiating a transition
    when appropriate, rather than abruptly walking away or rushing
    to the next location.

    \appendixblockseparator

    \item \textit{Bottleneck resolution.}
    When a plan was blocked by a missing character or a stalled
    situation, the focal agent(s) resolved the bottleneck in a socially
    believable and role-appropriate way, such as delegating a task or
    continuing to wait.

    \appendixblockseparator

    \item \textit{Group synchronization.} During shared
    activities or transitions, the focal agent(s) successfully
    synchronized with their companions, such as prompting them before
    leaving, rather than awkwardly abandoning or leaving them behind.

    \appendixblockseparator

    \item \textit{Overall social plausibility.} Overall, the
    focal agent(s) behaved believably as interactive and interdependent
    members of a Roman household while respecting the social norms and
    constraints of the setting.

\end{enumerate}


\appendixsubsectionheading{Statistical Analysis and Detailed Results}
\label{app:condition_comparison_statistics}

\paragraph{Analysis.}

\begin{figure}[t]
\centering
\includegraphics[width=\linewidth]{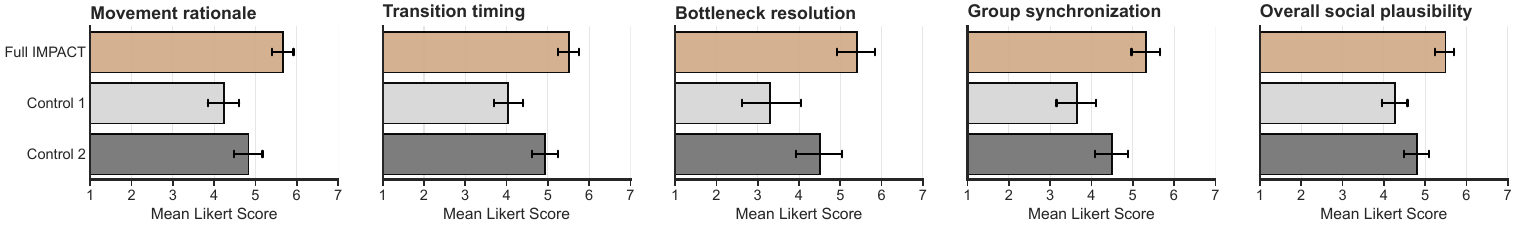}
\caption{Mean participant ratings of Full Architecture, Control~1, and
Control~2 across five measures. Ratings ranged from 1 to 7, with higher scores indicating
more favorable evaluations. Error bars represent 95\% confidence intervals.}
\label{fig:condition_ratings}
\par\smallskip
{\small\raggedright
\textbf{Alt text:} Five horizontal bar charts compare participant ratings of movement rationale, transition timing, bottleneck resolution, group synchronization, and overall social plausibility. Full Architecture has the highest mean rating on every measure, followed by Control 2 and Control 1.
\par}
\end{figure}

Because participants evaluated all three conditions but only a subset of events, the study followed a within-subjects design with incomplete event blocks. Participants could select \textit{Not applicable} when
a rating dimension did not meaningfully apply to a particular video.
These responses were excluded from the corresponding
dimension-specific analysis rather than assigned a numerical score.

For each dimension, we fitted a linear mixed-effects model with
condition and event as fixed effects and participant as a random
intercept. Full IMPACT was treatment-coded as the reference condition,
such that negative coefficients indicate lower ratings for a control
condition relative to Full IMPACT. We report fixed-effect coefficients
($b$), 95\% confidence intervals, and model-based pairwise comparisons.
Holm adjustment was applied to the three pairwise comparisons within
each dimension.

\appendixblockseparator

\paragraph{Valid observations.}

All 37 participants contributed ratings for movement rationale,
transition timing, and overall social plausibility. Because
bottleneck resolution and group synchronization applied only to
particular events, their available sample sizes were smaller. The
numbers of participants contributing ratings in each condition ranged
from 24 to 25 for bottleneck resolution and from 32 to 33 for group
synchronization. The models included 300 valid ratings for movement
rationale, 305 for transition timing, 75 for bottleneck resolution,
196 for group synchronization, and 329 for overall social
plausibility.

\appendixblockseparator

\paragraph{Condition effects.}

Full IMPACT received the highest observed mean rating on all five
dimensions (Supplementary Figure~\ref{fig:condition_ratings}). Relative to Full IMPACT, both control conditions received
lower model-estimated ratings on every dimension, with all corresponding
95\% confidence intervals excluding zero.

For \textit{movement rationale}, ratings were lower for Control~1
($b=-1.453$, 95\% CI $[-1.857,-1.048]$) and Control~2
($b=-0.858$, 95\% CI $[-1.263,-0.453]$). The same pattern occurred
for \textit{transition timing} (Control~1: $b=-1.494$, 95\% CI
$[-1.875,-1.113]$; Control~2: $b=-0.588$, 95\% CI
$[-0.969,-0.207]$), \textit{bottleneck resolution} (Control~1:
$b=-2.048$, 95\% CI $[-2.762,-1.334]$; Control~2: $b=-0.888$,
95\% CI $[-1.612,-0.163]$), \textit{group synchronization}
(Control~1: $b=-1.670$, 95\% CI $[-2.183,-1.157]$; Control~2:
$b=-0.840$, 95\% CI $[-1.354,-0.327]$), and \textit{overall
social plausibility} (Control~1: $b=-1.225$, 95\% CI
$[-1.564,-0.886]$; Control~2: $b=-0.702$, 95\% CI
$[-1.042,-0.362]$).

\appendixblockseparator

\paragraph{Pairwise comparisons.}

Holm-adjusted comparisons confirmed that Full IMPACT received higher
ratings than both controls on \textit{movement rationale}
(vs.\ Control~1: $p<.0001$; vs.\ Control~2: $p=.0001$),
\textit{transition timing} (vs.\ Control~1: $p<.0001$;
vs.\ Control~2: $p=.0025$), \textit{bottleneck resolution}
(vs.\ Control~1: $p<.0001$; vs.\ Control~2: $p=.0164$),
\textit{group synchronization} (vs.\ Control~1: $p<.0001$;
vs.\ Control~2: $p=.0027$), and \textit{overall social
plausibility} (vs.\ Control~1: $p<.0001$; vs.\ Control~2:
$p=.0001$).

Control~2 also received higher ratings than Control~1 on every
dimension: movement rationale ($p=.0040$), transition timing
($p<.0001$), bottleneck resolution ($p=.0033$), group
synchronization ($p=.0027$), and overall social plausibility
($p=.0026$). Thus, all three conditions differed significantly from
one another on each of the five dimensions, with the consistent
ordering Full IMPACT, Control~2, and Control~1 from highest to lowest
ratings.

\appendixsubsectionheading{Illustrative Behavioral Differences Across Conditions}
\label{app:condition_behavior_examples}

Supplementary Table~\ref{tab:condition_behavior_examples} summarizes the behaviors
observed under Full IMPACT, Control 1, and Control 2 for the
three events most frequently selected by participants in the video-based
evaluation. These examples illustrate how differences in social
milestone executability and wait-or-prompt resolution affected agents'
coordination and movement. The table was prepared for reporting
purposes and was not presented to participants during the study.

\begin{table*}[t]
\centering
\small
\caption{Observed behaviors under the three conditions for the events most
frequently selected by participants.}
\label{tab:condition_behavior_examples}
\begin{tabularx}{\textwidth}{
    >{\raggedright\arraybackslash}p{0.19\textwidth}
    >{\raggedright\arraybackslash}X
    >{\raggedright\arraybackslash}X
    >{\raggedright\arraybackslash}X}
\toprule
\textbf{Event} &
\textbf{Full IMPACT} &
\textbf{Control 1} &
\textbf{Control 2} \\
\midrule
\textbf{1. Servant Invites Guests to Wait for the Host} &
The servant waits for the guests, invites them, and accompanies them to the Reception Room. &
The servant greets the guests, and they proceed to the Reception Room in quick succession. &
The servant waits for the guests, invites them, and accompanies them to the Reception Room. \\

\addlinespace

\textbf{3. Host Greets the Guests in the Reception Room} &
The host waits for the servant's report before entering the Reception Room to greet the guests. &
The host enters the Reception Room before receiving the servant's report. &
The host seeks out the servant for the report and encounters the guests before they are settled. \\

\addlinespace

\textbf{8. Host Initiates the Dancer's Performance} &
Finding the dancer absent, the host sends a servant to summon her to the Dining Room. &
The performance phase begins despite the dancer's absence. &
The dancer arrives before being summoned, causing the performance phase to begin prematurely. \\

\bottomrule
\end{tabularx}
\end{table*}

\appendixsectionheading{Participant Evaluation Part 2: Module-Level Decision Evaluation}
\label{app:part_2_module_level}
\paragraph{Design and procedure.}
Part~2 assessed whether participants considered individual decisions generated by the Full IMPACT architecture appropriate and plausible. Following prior module-level evaluations of LLM-agent architectures~\cite{ren2024emergence}, we defined eight decision-item types covering the four modules: three for milestone planning (milestone generation, success-condition specification, and completion judgment), two for wait-or-prompt resolution (coordination with co-present agents and coordination when a required person was absent), one for directive issuance (directive expression), and two for directive integration (directive extraction and plan revision).

Each item presented the inputs relevant to the corresponding module, the LLM-generated output, and a question evaluating that output. For each of the eight item types, participants evaluated one randomly selected instance from the corresponding question pool, resulting in eight ratings per participant. 

Participants rated their agreement with each decision on a seven-point Likert scale ranging from 1 (\textit{strongly disagree}) to 7 (\textit{strongly agree}). 

\appendixsubsectionheading{Evaluation Questionnaire}
\label{app:decision_level_questionnaire}

This appendix presents the eight question formats used in Part~2 of
the human evaluation. Each participant evaluated one randomly sampled
decision points for each question type from the five complete runs. Placeholders indicate
information populated from the corresponding trace. 
We resampled the dataset to ensure that completed and incomplete milestone judgments were equally represented in the pool; waiting and prompt decisions were also equally
represented. Items evaluating directive issuance, identification, and
plan integration were sampled from interactions containing an
identifiable directive. 

All items used a seven-point agreement scale:

\begin{appendixlinedblock}
\begin{quote}
1 = Strongly disagree;
2 = Disagree;
3 = Somewhat disagree;
4 = Neutral;
5 = Somewhat agree;
6 = Agree;
7 = Strongly agree.
\end{quote}
\end{appendixlinedblock}

\subsubsection{Socially Gated Milestone Planning}
\label{app:questionnaire_m1}

\begin{questionnaireitem}{Question 1: Milestone Rationality}

\qfield{QUESTION}

Do you agree that the planned milestones accurately reflect the
expected presence of surrounding characters?

\qfield{AGENT NAME AND ROLE}

\placeholder{agent name and social role}

\qfield{PREDEFINED TIMELINE}

\placeholder{agent's predefined timeline of the dinner event}

\qfield{GENERATED MILESTONES}

Milestone 1:
\placeholder{activity goal}

Planned location:
\placeholder{location}

Required participants:
\placeholder{agents required to complete the milestone}

\smallskip

Milestone 2:
\placeholder{activity goal}

Planned location:
\placeholder{location}

Required participants:
\placeholder{agents required to complete the milestone}

\smallskip

\placeholder{additional generated milestones}

\end{questionnaireitem}

\begin{questionnaireitem}{Question 2: Milestone Transition Condition}

\qfield{QUESTION}

Is this trigger logical and sufficient to allow
\placeholder{agent name} to move to the next activity?

\qfield{AGENT NAME AND ROLE}

\placeholder{agent name and social role}

\qfield{CURRENT ACTIVITY}

Location:
\placeholder{current milestone location}

Goal:
\placeholder{current milestone goal}

Required participants:
\placeholder{social configuration required for the current activity}

\qfield{NEXT ACTIVITY}

Location:
\placeholder{next milestone location}

Goal:
\placeholder{next milestone goal}

Required participants:
\placeholder{social configuration required for the next activity}

\qfield{TRIGGER OPTIONS}

1.
\placeholder{socially dependent transition condition}

2.
\placeholder{completion of the current milestone's detailed actions}

\qfield{GENERATED SUCCESS TRIGGER}

\placeholder{selected milestone completion condition}

\end{questionnaireitem}

\begin{questionnaireitem}{Question 3: Milestone Completion Judgment}

\qfield{QUESTION}

Based on the current observation, do you agree with the LLM's
judgment on whether the milestone is finished?

\qfield{AGENT NAME AND ROLE}

\placeholder{agent name and social role}

\qfield{CURRENT MILESTONE}

\placeholder{current milestone goal}

\qfield{SUCCESS TRIGGER}

\placeholder{observable condition required to complete the milestone}

\qfield{CURRENT OBSERVATION}

\placeholder{agent's current location}

\placeholder{observed agents and their current actions}

\placeholder{relevant arrivals, departures, or conversations}

\qfield{AGENT'S JUDGMENT}

\placeholder{finished or not finished}

\qfield{REASONING}

\placeholder{explanation of the milestone-completion judgment}

\end{questionnaireitem}

\subsubsection{Wait-or-Prompt Resolution}
\label{app:questionnaire_m2}

\begin{questionnaireitem}{Question 1: Movement Coordination Strategy}

\qfield{QUESTION}

Do you agree that the agent's chosen action strategy is appropriate
and believable given their social role and the current situation?

\qfield{AGENT NAME AND ROLE}

\placeholder{agent name and social role}

\qfield{CURRENT MILESTONE}

Goal:
\placeholder{current milestone goal}

Location:
\placeholder{current milestone location}

Success trigger:
\placeholder{condition required to complete the current milestone}

\qfield{NEXT PLANNED MILESTONE}

Goal:
\placeholder{next milestone goal}

Location:
\placeholder{next milestone location}

\qfield{CURRENT CONTEXT}

\placeholder{agent's current location}

\placeholder{other agents present and their ongoing activities}

\qfield{AVAILABLE STRATEGIES}

\textbf{WAIT:}
Remain in the current location instead of initiating movement.

\textbf{INVITE CO-MOVEMENT:}
Invite other agents to move together to the next location.

\textbf{INFORM MOVING:}
Inform other agents of an upcoming departure.

\qfield{SELECTED STRATEGY}

\placeholder{selected coordination strategy}

\end{questionnaireitem}

\begin{questionnaireitem}{Question 2: Missing-Participant Resolution}

\qfield{QUESTION}

Do you agree that the agent's chosen action strategy is appropriate
and believable given their social role and the current situation?

\qfield{AGENT NAME AND ROLE}

\placeholder{agent name and social role}

\qfield{CURRENT MILESTONE}

Goal:
\placeholder{current milestone goal}

Success trigger:
\placeholder{condition involving a currently absent person}

\qfield{NEXT PLANNED MILESTONE}

Goal:
\placeholder{next milestone goal}

Location:
\placeholder{next milestone location}

\qfield{CURRENT CONTEXT}

\placeholder{agent's current location}

\placeholder{relevant agents present or absent}

\qfield{AVAILABLE STRATEGIES}

\textbf{WAIT:}
Remain in the current location and continue waiting.

\textbf{DELEGATE FIND AND DIRECT:}
Ask another agent to locate the missing person and deliver a
directive.

\textbf{FIND AND DIRECT PERSONALLY:}
Personally locate the missing person and deliver a directive.

\qfield{SELECTED STRATEGY}

\placeholder{selected coordination strategy}

\end{questionnaireitem}

\subsubsection{Structured Directive Issuance}
\label{app:questionnaire_m3}

\begin{questionnaireitem}{Question 1: Directive Formulation}

\qfield{QUESTION}

Do you agree that the agent clearly conveyed the intended social
directive to the conversational partner?

\qfield{AGENT NAME AND ROLE}

\placeholder{sender's name and social role}

\qfield{CONVERSATIONAL PARTNER NAME AND ROLE}

\placeholder{recipient names and social roles}

\qfield{CURRENT CONTEXT}

\placeholder{sender's ongoing activity}

\placeholder{recipient activities and conversational context}

\qfield{SOCIAL DIRECTIVE INTENTION}

Directive type:
\placeholder{invitation or request}

Recipients:
\placeholder{intended recipients}

Target location:
\placeholder{location associated with the directive}

Requested action:
\placeholder{intended recipient action}

\qfield{GENERATED CONVERSATION}

\placeholder{sender's first utterance}

\placeholder{recipient's response}

\placeholder{additional conversational exchanges}

\end{questionnaireitem}

\subsubsection{Directive Integration}
\label{app:questionnaire_m4}

\begin{questionnaireitem}{Question 1: Directive Identification}

\qfield{QUESTION}

Do you agree that the receiver accurately extracted the correct
social directive from the conversation transcript?

\qfield{SENDER AND RECEIVER}

\placeholder{sender name and social role}
$\rightarrow$
\placeholder{receiver name and social role}

\qfield{CONVERSATION TRANSCRIPT}

\placeholder{sender's directive-related utterance}

\placeholder{recipient's response}

\placeholder{additional conversational exchanges}

\qfield{EXTRACTED DIRECTIVE}

Directive type:
\placeholder{request or invitation}

Target location:
\placeholder{identified location}

Requested action:
\placeholder{identified action}

\end{questionnaireitem}

\begin{questionnaireitem}{Question 2: Directive Integration}

\qfield{QUESTION}

Do you agree that
\placeholder{agent name}
appropriately incorporated the received social directive into a new
milestone, including its goal, estimated duration, and success
condition?

\qfield{AGENT NAME AND ROLE}

\placeholder{recipient's name and social role}

\qfield{PREVIOUS MILESTONE PLAN}

Goal:
\placeholder{previous milestone goal}

Location:
\placeholder{previous milestone location}

Required participants:
\placeholder{previously required social configuration}

Estimated duration:
\placeholder{previous milestone duration}

\qfield{RECEIVED DIRECTIVE}

Sender:
\placeholder{directive sender}

Directive type:
\placeholder{invitation or request}

Requested action:
\placeholder{requested activity}

Target location:
\placeholder{requested destination}

\qfield{UPDATED MILESTONE PLAN}

Goal:
\placeholder{revised milestone goal}

Location:
\placeholder{revised milestone location}

Required participants:
\placeholder{updated required social configuration}

Estimated duration:
\placeholder{estimated duration of the revised milestone}

Success Condition (condition indicates completion of the milestone plan):
\placeholder{condition required to complete the revised milestone}

\end{questionnaireitem}

\paragraph{Results.}

\begin{table}[t]
\centering
\caption{Participant ratings of eight sampled decisions from the four IMPACT modules. Scores report the mean and sample standard deviation on a seven-point agreement scale.}
\label{tab:module_scores}

\scriptsize
\renewcommand{\arraystretch}{1.5}
\setlength{\tabcolsep}{1.3pt}
\setlength{\extrarowheight}{3pt} 
\setlength{\arrayrulewidth}{0.4pt}

\begin{tabular}{
@{}
>{\centering\arraybackslash}m{0.14\columnwidth} |
>{\centering\arraybackslash}m{0.19\columnwidth}
>{\centering\arraybackslash}m{0.13\columnwidth} |
>{\centering\arraybackslash}m{0.14\columnwidth} |
>{\centering\arraybackslash}m{0.19\columnwidth}
>{\centering\arraybackslash}m{0.13\columnwidth}
@{}
}

\hline

\textbf{Module} &
\shortstack{\textbf{Sub-}\\[-1pt]\textbf{component}} &
\textbf{Score} &
\textbf{Module} &
\shortstack{\textbf{Sub-}\\[-1pt]\textbf{component}} &
\textbf{Score} \\

\hhline{-|--|-|--}

\multirow{3}{*}{\shortstack{Milestone\\[-1pt]Planning}} &
\shortstack{Milestone\\[-1pt]Generation} &
\shortstack{$5.81$\\[-1pt]$\pm 0.78$} &
\multirow{2}{*}{\shortstack{Wait-or-\\[-1pt]Prompt}} &
\shortstack{Movement\\[-1pt]Coordination} &
\shortstack{$5.84$\\[-1pt]$\pm 1.09$} \\

\hhline{~|--|~|--}

&
\shortstack{Success\\[-1pt]Conditions} &
\shortstack{$5.35$\\[-1pt]$\pm 1.24$} &
&
\shortstack{Missing Person\\[-1pt]Coordination} &
\shortstack{$5.62$\\[-1pt]$\pm 1.40$} \\

\hhline{~|--|-|--}

&
\shortstack{Completion\\[-1pt]Judgment} &
\shortstack{$5.14$\\[-1pt]$\pm 1.75$} &
\multirow{2}{*}{\shortstack{Directive\\[-1pt]Integration}} &
\shortstack{Directive\\[-1pt]Extraction} &
\shortstack{$5.95$\\[-1pt]$\pm 1.43$} \\

\hhline{-|--|~|--}

\shortstack{Directive\\[-1pt]Issuance} &
\shortstack{Directive\\[-1pt]Expression} &
\shortstack{$5.95$\\[-1pt]$\pm 1.20$} &
&
\shortstack{Plan\\[-1pt]Revision} &
\shortstack{$5.14$\\[-1pt]$\pm 1.44$} \\

\hline

\end{tabular}
\end{table}
Supplementary Table~\ref{tab:module_scores} summarizes the decision-level ratings. All eight item types received positive mean ratings above the neutral midpoint, ranging from $5.14$ to $5.95$. Directive Extraction ($M=5.95$, $SD=1.43$), and Directive Expression ($M=5.95$, $SD=1.20$) received the highest rating. These results indicate that participants generally considered directives to be clearly communicated by senders and correctly recovered by recipients. Movement Coordination ($M=5.84$, $SD=1.09$) and Milestone Generation ($M=5.81$, $SD=0.78$) were also rated favorably.

Completion Judgment and Plan Revision received the lowest mean ratings (both $M=5.14$), although both remained above the neutral midpoint. Completion Judgment also showed the greatest variability ($SD=1.75$). These results suggest less agreement about two transition points in the architecture: determining when a milestone has been completed and incorporating a received directive into an existing plan. Overall, the evaluation supports the reasonableness of decisions produced across all four modules while identifying milestone completion and plan revision as areas for further refinement.

\appendixsectionheading{Semi-Structured Interview Guide}
\label{app:expert_interview_guide}

The following semi-structured interview guide was used to examine
participants' perspectives on the simulation's historical plausibility,
limitations, and potential applications. Follow-up questions were used
as appropriate to clarify or expand on participants' responses.

\appendixsubsectionheading{What Worked Well}

\noindent\textbf{Primary question}

\smallskip
\noindent
Thinking about the AI agents' movement, what, if anything, do you feel
worked particularly well or felt historically plausible?

\medskip
\noindent\textbf{Follow-up probes}

\begin{itemize}
    \item Can you walk me through a specific example of that?
    \item What specifically made that moment stand out as successful
    to you?
\end{itemize}

\appendixsubsectionheading{What Needs Improvement}

\noindent\textbf{Primary question}

\smallskip
\noindent
On the flip side, what aspects of the simulation, if any, do you feel
need the most work or felt out of character for the historical context?

\medskip
\noindent\textbf{Follow-up probes}

\begin{itemize}
    \item Could you explain why that interaction felt unrealistic?
    \item How would you suggest we adjust or improve it to make it
    more credible?
\end{itemize}

\appendixsubsectionheading{Value and Applications}

\noindent\textbf{Value for experts}

\smallskip
\noindent
Based on what you've seen, what value or practical applications do you
see this type of simulation having for your own learning or research,
or for others with your level of expertise in archaeology and history?

\medskip
\noindent\textbf{Value for broader audiences}

\smallskip
\noindent
How do you think that changes for a broader audience? For example,
what value might this have for students or the general public who
don't have a deep background in Roman history?

\medskip
\noindent\textbf{Comparison with existing tools or experiences}

\smallskip
\noindent
If you had to compare using or observing this Pompeii simulation to an
existing tool, experience, or concept, what would it be and why?

\medskip
\noindent\textbf{Understanding ancient behavior}

\smallskip
\noindent
How do you think simulations like this could help researchers to
understand ancient people's behavior?

\medskip
\noindent\textbf{Limitations}

\smallskip
\noindent
What do you think are the limitations that must be addressed for
historians to adopt it in this way?

\appendixsubsectionheading{Closing Question}

\noindent
That covers the main topics I wanted to discuss. Before we wrap up,
is there anything else you noticed, any wildcard thoughts, or any
questions you have for me that we haven't touched on yet?



\end{appendices}

\end{document}